\documentclass [11pt] {article}
\RequirePackage{algorithm,algorithmic,amssymb,epsf,epic}

\def\denseformat{
\setlength{\textheight}{9in}
\setlength{\textwidth}{6.9in}
\setlength{\evensidemargin}{-0.2in}
\setlength{\oddsidemargin}{-0.2in}
\setlength{\headsep}{10pt}
\setlength{\topmargin}{-0.3in}
\setlength{\columnsep}{0.375in}
\setlength{\itemsep}{0pt}
}

\newtheorem{theorem}{Theorem}[section]

\newtheorem{claim}[theorem]{Claim}
\newtheorem{lemma}[theorem]{Lemma}

\newtheorem{corollary}[theorem]{Corollary}
\newtheorem{fact}[theorem]{Fact}
\newtheorem{comment}[theorem]{Comment}

\def\boldhead#1:{\par\vskip 7pt\noindent{\bf #1:}\hskip 10pt}
\def\ithead#1:{\par\vskip 7pt\noindent{\it #1:}\hskip 10pt}

\def\inline#1:{\par\vskip 7pt\noindent{\bf #1:}\hskip 10pt}
\def\midinline#1:{\par\noindent{\bf #1:}\hskip 10pt}
\def\dnsinline#1:{\par\vskip -7pt\noindent{\bf #1:}\hskip 10pt}
\def\ddnsinline#1:{\newline{\bf #1:}\hskip 10pt}
\def\largeinline#1:{\par\vskip 7pt\noindent{\large\bf #1:}\hskip 10pt}
\long\def\comment #1\commentend{}
\long\def\commhide #1\commhideend{}
\long\def\commfull #1\commend{#1}
\long\def\commabs #1\commenda{}
\long\def\commtim #1\commendt{#1}
\long\def\commb #1\commbend{}
\long\def\commedit #1\commeditend{} 

\long\def\commB #1\commBend{}       

\long\def\commex #1\commexend{}     

\long\def\commsiena #1\commsienaend{}  
                                         
\long\def\commBI #1\commBIend{}  
                                         
\long\def\CProof #1\CQED{}

\def\blackslug{\hbox{\hskip 1pt \vrule width 4pt height 8pt
    depth 1.5pt \hskip 1pt}}
\def\QED{\quad\blackslug\lower 8.5pt\null\par}

\def\Proof{\par\noindent{\bf Proof:~}}

\def\proof{\Proof}

\long\def\PPP#1{\noindent{\bf Proof:}{ #1}{\quad\blackslug\lower 8.5pt\null}}
\long\def\PP#1{\noindent {\bf Proof}:{ #1} $\Box$ \\}

\long\def\denspar #1\densend
{#1}

\newif\ifnotesw\noteswtrue
   {\ifnotesw\marginpar[\hfill\(\top\)]{\(\top\)}\fi}%
   {\ifnotesw\marginpar[\hfill\(\bot\)]{\(\bot\)}\fi}

\newcommand{\mnote}[1]%
    {\ifnotesw\marginpar%
        [{\scriptsize\it\begin{minipage}[t]{\marginparwidth}
        \raggedleft#1%
                        \end{minipage}}]%
        {\scriptsize\it\begin{minipage}[t]{\marginparwidth}
        \raggedright#1%
                        \end{minipage}}%
    \fi}

\def\MathF{\hbox{\rm I\kern-2pt F}}
\def\MathP{\hbox{\rm I\kern-2pt P}}
\def\MathR{\hbox{\rm I\kern-2pt R}}
\def\MathZ{\hbox{\sf Z\kern-4pt Z}}
\def\MathN{\hbox{\rm I\kern-2pt I\kern-3.1pt N}}
\def\MathC{\hbox{\rm \kern0.7pt\raise0.8pt\hbox{\footnotesize I}
\kern-4.2pt C}}
\def\MathQ{\hbox{\rm I\kern-6pt Q}}

\newsavebox{\ttop}\newsavebox{\bbot}

\def\polylog{\mbox{polylog}}

\denseformat

\begin{document}

\newcommand {\ignore} [1] {}

\def \bfm {\bf \boldmath}

\def \AA {{\alpha}}
\def \BB {{\beta}}
\def \CC {{\gamma}}
\def \PP {{\cal P}}

\def \pa {{\boldmath $A$}}
\def \pb {{\boldmath $B$}}
\def \pc {{\boldmath $C$}}
\def \px {{\boldmath $X$}}
\def \py {{\boldmath $Y$}}
\def \pz {{\boldmath $Z$}}

\title{Shallow, Low, and Light Trees, and\\ Tight Lower Bounds
for Euclidean Spanners
}

\author{
Yefim Dinitz \thanks{
        Department of Computer Science,
        Ben-Gurion University of the Negev,
        POB 653, Beer-Sheva 84105, Israel.
        E-mail: {\tt \{dinitz,elkinm,shayso\}@cs.bgu.ac.il}
        \newline Partially supported by the Lynn and William Frankel
        Center for Computer Sciences.}
         \and
Michael Elkin$^*$\thanks{This research has been supported by the
Israeli Academy of Science, grant 483/06.}
         \and
Shay Solomon${^*}{^\dagger}$ }
\date{\empty}

\date{\empty}

\begin{titlepage}
\def\thepage{}
\maketitle

\begin {abstract}
  We show that for every $n$-point metric space $M$ there
  exists a spanning tree $T$ with
  unweighted diameter  $O(\log n)$
  and weight $\omega(T) = O(\log n) \cdot \omega(MST(M))$.
  Moreover, there is a designated point $rt$  such that for
  every point $v$, $dist_T(rt,v) \le (1+\epsilon) \cdot
  dist_M(rt,v)$, for an arbitrarily small constant $\epsilon > 0$.
  We extend this result, and provide a tradeoff between
  unweighted diameter and weight, and prove that this tradeoff is
  \emph{tight up to constant factors} in the entire range of
  parameters.

  These results enable us to settle
  a long-standing open question in Computational Geometry.
  In STOC'95 Arya et al. devised a construction of Euclidean
  Spanners with unweighted diameter $O(\log n)$ and weight
   $O(\log n) \cdot \omega(MST(M))$. Ten years later in SODA'05
   Agarwal et al. showed that this result is tight up to a factor of
   $O(\log \log n)$. We close this gap and show that the result of
   Arya et al. is tight up to constant factors.
\end{abstract}
\end{titlepage}

\pagenumbering {arabic} 


\section{Introduction} \label{s:intro}
\subsection{Background and Main Results}

Spanning trees for finite metric spaces have been a subject of an
ongoing intensive research since the beginning of the nineties
\cite{AKPW95,Bar96,Bar98,CCGGP98,FRT03,EEST05,BKJ83,Jaf85,ABP90,Peleg00,ABP91,RSMRR94}.
In particular, many researchers studied the notion of
\emph{shallow-light trees}, henceforth SLT
\cite{BKJ83,Jaf85,ABP90,ABP91,RSMRR94,AHHKK95,Peleg00}. Roughly
speaking, SLT of an $n$-point metric space $M$ is a spanning tree
$T$ of the complete graph corresponding to $M$ whose total weight is
close to the weight $w(MST(M))$ of the minimum spanning tree
$MST(M)$ of $M$, and whose weighted diameter is close to that of
$M$. (See Section \ref{sec:prel} for relevant definitions.)

In addition to being an appealing combinatorial object, SLTs turned
out to be useful for various data gathering and dissemination
problems in the message-passing model of distributed computing
\cite{ABP90}, in approximation algorithms \cite{RSMRR94}, for
constructing spanners \cite{ABP91,AHHKK95}, and for VLSI-circuit
design \cite{CKRSW91,CKRSW92,CKRSW292}. Near-optimal tradeoffs
between the weight and diameter of SLTs were established by Khuller
et al. \cite{KRY94}, and by Awerbuch et al. \cite{ABP91}.

Even though the requirement that the spanning tree $T$ will have a
small weighted-diameter is a natural one, it is no less natural to
require it to have a small \emph{unweighted diameter} (also called
\emph{hop-diameter}). The latter requirement guarantees that any two
points of the metric space will be connected in $T$ by a path that
consists of only a small \emph{number of edges} or \emph{hops}. This
guarantee turns out to be particularly important for routing
\cite{HP00,AM04}, computing almost shortest paths in sequential and
parallel setting \cite{Coh93,Coh94,DHZ96}, and in other
applications. Another parameter that plays an important role in many
applications is the maximum (vertex) degree of the constructed tree
\cite{ADMSS95,BGS05,AS97,HP00}.

In this paper we introduce and investigate a related notion of
\emph{low-light trees}, henceforth LLTs, that combine small weight
with small hop-diameter. We present near-tight upper and lower
bounds on the parameters of LLTs. In addition, our constructions of
LLTs have {\em optimal maximum degree}.

To specify our results, we need some notation. For a spanning tree
$T$ of a metric $M$, let $\Lambda = \Lambda(T)$ denote the
hop-diameter of $T$, and $\Psi = \Psi(T) =
\frac{\omega(T)}{\omega(MST(M))}$ denote the ratio between its
weight and the weight of the minimum spanning tree of $M$,
henceforth the \emph{lightness} of $T$. In particular, we show the
following bounds that are tight up to constant factors in the
\emph{entire} range of the parameters.

\begin{figure*}[htp]
\begin{center}
\begin{minipage}{\textwidth} 
\begin{center}
\setlength{\epsfxsize}{1.5in}
\epsfbox{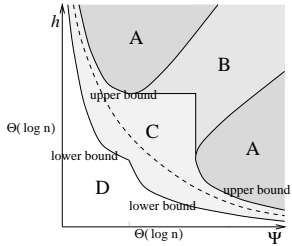}
\end{center}
\end{minipage}
\caption[]{ \label{fig:bnd} \sf The dashed line separates two sets
of pairs $(\Psi,h)$. For a pair $(\Psi,h)$ above the line, for any
$n$-point metric space there exists a spanning tree with lightness
at most $\Psi$ and hop-radius at most $h$. For a pair $(\Psi,h)$
below the line, there exist $n$-point metric spaces for which this
property does not hold. The two areas A and the area B are all
contained in the former set, while the area D is contained in the
latter one. The two areas A depict our upper bound constructions,
and their extension by monotonicity is depicted by the area B. The
area $D$ represents our lower bounds. The area $C$ represents the
gap between our upper and lower bounds. }
\end{center}
\end{figure*}

\begin {enumerate}
\item For any sufficiently large integer $n$ and positive integer $h$,
and an $n$-point metric space $M$, there exists a spanning tree of
$M$ with \emph{hop-radius}\footnote{\emph{Hop-radius} $h(G,rt)$ of a
graph $G$ with respect to a distinguished vertex $rt$ is the maximum
number of hops in a simple path connecting the vertex $rt$ with some
vertex $v$ in $G$. Obviously, $h(G,rt) \le \Lambda(G) \le 2 \cdot
h(G,rt)$. For a rooted tree $(T,rt)$, the hop-radius (called also
{\em depth}) of $(T,rt)$ is the hop-radius of $G$ with respect to
$rt$. Hop-radius of $G$, denoted $h(G)$, is defined by $h(G) = \min
\{ h(G,rt) \mid rt \in V\}$. }
 at most $h$ and lightness at most $O(\Psi)$, for $\Psi$ that satisfies
the following relationship. If $h \ge \log n$ then ($\Psi$ is at
most $O(\log n)$ and $h = O(\Psi \cdot n^{1/\Psi})$). In the complementary range $h < \log n$,
it holds that $\Psi=O(h\cdot n^{1/h})$.
\\Moreover, this spanning tree is a binary one whenever $h \ge \log
n$, and it has the {\em optimal} maximum degree $\left \lceil n^{1/h}
\right \rceil$ whenever $h < \log n$.
In addition, in the
entire range of parameters the respective spanning trees can be
constructed in polynomial time.
\item For $n$ and $h$ as above, and $h \ge \log n$,
there exists an $n$-point metric space $M^*=M^*(n)$ for which any
spanning subgraph with hop-radius at most $h$ has lightness at least
$\Omega(\Psi)$, for some $\Psi$ satisfying $h = \Omega(\Psi \cdot
n^{1/\Psi})$.
\item For $n$ and $h$ as above, and $h < \log n$, any spanning subgraph with
hop-radius at most $h$ for $M^*(n)$ has lightness at least $\Psi =
\Omega(h \cdot n^{1/h})$.
\end {enumerate}
(Note that the equation $x \cdot n^{1/x} = \Theta(\log n)$
holds if and only if $x = \Theta(\log n)$.)
 See Figure \ref{fig:bnd} for an illustration of our results.

The small maximum degree of our LLTs may be helpful for various
applications in which the degree of a vertex $v$ corresponds to the
load on a processor that is located in $v$. The requirement to
achieve small maximum degree is particularly important for
applications in Computational Geometry. (See
\cite{ADMSS95,BGS05,AS97}, and the references therein.)

\subsection{Lower Bounds for Euclidean Spanners}
While our upper bounds apply to all finite metric spaces, our lower
bounds apply to an extremely basic metric space $M^* = \vartheta_n$.
Specifically, this metric space is the 1-dimensional Euclidean space
with $n$ points $v_1,v_2,\ldots,v_n$ lying on the $x$-axis with
coordinates $1,2,\ldots,n$, respectively. The basic nature of
$\vartheta_n$ strengthens our lower bounds, as they are applicable
even for very limited classes of metric spaces. One particularly
important application of our lower bounds is in the area of
Euclidean Spanners. For a set $\mathcal U$ of $n$ points in $\mathbb
R^2$, and a parameter $\alpha$, $\alpha \ge 1$, a subset $\mathcal
H$ of the ${n \choose 2}$ segments connecting pairs of points from
$\mathcal U$ is called an (Euclidean) \emph{$\alpha$-spanner} for
$\mathcal U$, if for every pair of points $u,v \in \mathcal U$, the
distance between them in $\mathcal H$ is at most $\alpha$ times the
Euclidean distance between them in the plane. Euclidean spanner is a
very fundamental geometric construct with numerous applications in
Computational Geometry  \cite{ADMSS95,AMS94,AS97} and Network Design
\cite{HP00,MP00}. (See the recent book of Narasimhan and Smid
\cite{NS07} for a detailed account on Euclidean spanners and their
applications.)

A seminal paper that was a culmination of a long line of research on
Euclidean spanners was published by Arya et al. \cite{ADMSS95} in
STOC'95. One of the main results of this paper is a construction of
$(1+\epsilon)$-spanners with $O(n)$ edges that also have lightness
and hop-diameter both bounded by $O(\log n)$. As an evidence of the
optimality of this combination of parameters, Arya et al. cited a
result by Lenhof et al. \cite{LSW94}. Lenhof et al. showed that any
construction of Euclidean spanners that employs well-separated pair
decompositions cannot achieve a better combination of weight and
hop-diameter. However, the fundamental question of whether this
combination of parameters can be improved by other means was left
open in Arya et al. \cite{ADMSS95}. A partial answer to this
intriguing problem was given by Agarwal et al. \cite{AWY05} in
SODA'05. Specifically, it is shown in \cite{AWY05} that any
Euclidean spanner with lightness $O(\log n)$ must have diameter at
least $\Omega(\frac{\log n}{\log \log n})$, and vice versa.
Consequently, Agarwal et al. showed that the upper bound of Arya et
al. is optimal up to a factor $O(\log \log n)$. A simple corollary
of our lower bounds is that the result of Arya et al. is tight up to
constants even for \emph{one-dimensional} spanners! In other words,
we show that if the lightness is $O(\log n)$ then the diameter is
$\Omega(\log n)$ and vice versa, settling the open problem of
\cite{ADMSS95,AWY05}.

\subsection{Shallow-Low-Light-Trees}
We show that our constructions of LLTs extend to provide also a good
approximation of all \emph{weighted} distances from any given
designated root vertex $rt$. The resulting spanning trees achieve
small weight, hop-diameter, and weighted-diameter
\emph{simultaneously}! In other words, these trees combine the
useful properties of SLTs and LLTs in \emph{one construction}, and
thus we call them \emph{shallow-low-light-trees}, henceforth SLLTs.

Specifically, we show that for any sufficiently large integer
$n$, a positive integer $h$, a positive real $\epsilon > 0$, an
$n$-point metric space $M$, and a designated root point $rt$,
there exists a spanning tree $T$ of $M$ rooted at $rt$ with
hop-radius at most $O(h)$ and lightness at most $O(\Psi \cdot
(\epsilon^{-1}))$, such that ($h = O(\Psi \cdot n^{1/\Psi})$ and
$\Psi = O(\log n)$) whenever $h \ge \log n$, and $\Psi = O(h \cdot
n^{1/h})$ whenever $h < \log n$. Moreover, for every point $v \in M$,
the weighted distance between the root $rt$ and $v$ in $T$ is
greater by at most a factor of $(1+\epsilon)$ than the weighted
distance between them in $M$. This combination of parameters is
optimal up to constant factors. Finally, all our constructions
can be implemented in polynomial time.

We believe that this construction may be particularly useful in
algorithmic applications. In particular, Awerbuch et al.\
\cite{ABP90} presented the notion of cost-sensitive communication
complexity to the analysis of distributed algorithms. They used SLTs
to devise efficient algorithms with respect to the cost-sensitive
communication complexity for a plethora of basic problems in the
area of Distributed Computing, including network synchronization,
global function computation, and controller protocols. However,
their algorithms may perform quite poorly with respect to the
standard (not cost-sensitive) communication complexity notion. If
one could use SLLTs instead of SLTs in the construction of
\cite{ABP91}, it would result in distributed algorithms that are
efficient with respect to both standard and cost-sensitive notions
of communication complexity.

A major difficulty in implementing this scheme is that the
construction  of Awerbuch et al.\ \cite{ABP91} provides an SLT which
uses only edges of the original network, while our construction of
SLLTs applies to metric spaces, and thus it may employ edges that
are not present in the original network. Moreover, we show (see
Section \ref{sec:sllt}) that there are graphs with constant
hop-diameter for which any spanning tree has either huge
hop-diameter or huge weight, and thus there is no hope that LLTs or
SLLTs for general networks will be ever constructed. However, this
approach seems to be applicable for distributed algorithms that run
in \emph{complete} networks\footnote{Complete network is a network
in which every pair of processors is connected by a direct link.}
\cite{Sin92,LPP01}, \emph{overlay} networks \cite{DKKP95, AAABMP03},
and in other network architectures in which either direct or virtual
link may be readily established between each pair of processors.

To summarize, the problem of understanding the inherent tradeoff
between different parameters of LLTs is a fundamental one in the
investigation of spanning trees for metric spaces and graphs. In
addition, this basic and combinatorially appealing problem has
important applications to Computational Geometry and Distributed
Computing. We believe that further investigation of LLTs will expose
their additional applications, and connections to other areas.

 \subsection {Overview and Our Techniques}
   The most technically challenging
    part of our proof is the lower bound
for the range of $\Lambda \ge \log n$. The proof of this lower bound
consists of a number of components.
 First, we restrict our attention to
binary trees. Second, we adapt a linear program for the minimum
linear arrangement problem from the seminal paper of Even, Naor, Rao
and Schieber \cite{ENRS95} on spreading metrics to our needs. Third,
we analyze this linear program and show that the problem of
providing a lower bound for its solution reduces to a clean
combinatorial problem, and solve this problem. This enables us to
establish the desired lower bounds for \emph{binary trees}. Finally,
we extend those lower bounds to general trees by demonstrating that
our problem on general trees reduces to the same problem restricted
to binary trees.

The proof of our lower bounds for $\Lambda < \log n$ combines some
ideas from Agarwal et al. \cite{AWY05} with numerous new ideas.
Specifically, Agarwal et al. reduce the problem from the general
family of spanning subgraphs for $\vartheta_n$ to a certain
restricted family of \emph{stack graphs}. This reduction of
\cite{AWY05} provides a very elegant way for achieving somewhat
weaker bounds, but it is inherently suboptimal. In our proof we
tackle the general family of graphs directly. This more direct
approach results in  a much more technically involved proof, and in
much more accurate bounds.

For upper bounds we essentially reduce the problem of constructing
LLTs for general metric spaces to the same problem on $\vartheta_n$.
Somewhat surprisingly, despite the apparent simplicity of the metric
space $\vartheta_n$, the problem of constructing LLTs for this space
appears to be quite complex.

\subsection {Related work}
SLTs were extensively studied for the last twenty years
\cite{BKJ83,Jaf85,ABP90,ABP91,CKRSW91,CKRSW92,CKRSW292,KRY94,AHHKK95}.
However, all these constructions of SLT may result in trees with
very large hop-diameter, and the techniques used in those
constructions appear to be inapplicable to the problem of
constructing LLTs.

Euclidean spanners are also a subject of a recent extensive and
intensive research (see \cite{ADMSS95,DN94,AWY05,AMS94}, and the
references therein). However, the basic technique for constructing
them relies heavily on the methodology of well-separated pair
decomposition due to Callahan and Kosaraju \cite{CK92}. This
extremely powerful methodology is, however, applicable only for the
Euclidean metric space of constant dimension, while our
constructions apply to general metric spaces. Tight lower bounds on
the hop-diameter of Euclidean spanners with a given number of edges
were recently established by Chan and Gupta \cite{CG06}.
Specifically, it is shown in \cite{CG06} that for any $\epsilon > 0$
there exists an $n$-point Euclidean metric space $M = M(n,\epsilon)$
for which any Euclidean $(1+\epsilon)$-spanner with $m$ edges has
hop-diameter $\Omega(\alpha(m,n))$, where $\alpha$ is the functional
inverse of the Ackermann's function. Moreover, the metric space $M$
is 1-dimensional. (On the other hand, the space $M$ is still not as
restricted as $\vartheta_n$.) However, this lower bound provides no
indication whatsoever as to how \emph{light} can be Euclidean
spanners with low hop-diameter. In particular, the construction of
Arya et al. \cite{ADMSS95} that provides matching upper bounds to
the lower bounds of \cite{CG06} produces spanners that may have very
large weight.

In terms of the techniques, Chan and Gupta \cite{CG06} start with
showing their lower bounds for metrics induced by binary
hierarchically-separated-trees (henceforth, HSTs), and then
translate them into lower bounds for metrics induced by $n$ points
on the real line using known results. Their proof of the lower bound
for HSTs is an extension of Yao's proof technique from \cite{Yao82}.
As was discussed above, our lower bounds are achieved by completely
different proof techniques that involve analyzing a linear program
for the minimum linear arrangement problem. In particular, our lower
bounds are proved directly for $\vartheta_n$.

The study of spanning trees of the 1-dimensional metric space
$\vartheta_n$ is related to the extremely well-studied problem of
computing partial-sums. (See the papers of Yao \cite{Yao82},
Chazelle and Rosenberg \cite{CR91}, ~P\u{a}tra\c{s}cu and Demaine
\cite{PD04}, and the references therein.) For a discussion about the
relationship between these two problems we refer to the introduction
of \cite{AWY05}.

The linear program for the minimum linear arrangement problem that
we use for our lower bounds was studied in \cite{ENRS95,RR98}. There
is an extensive literature on the minimum linear arrangement problem
itself \cite{CHKR06,FL07}.

Finally, the extension of our construction of LLTs to SLLTs is
achieved by employing the construction of SLTs of \cite{ABP91} on
top of our construction of LLTs.

\subsection {The Structure of the Paper}
In Section \ref{sec:prel} we define the basic notions and present
the notation that is used throughout the paper. In Section
\ref{secmon} we show that the covering and weight functions, defined
in Section \ref{sec:prel}, are monotone non-increasing with the
depth parameter. This property is employed in Sections
\ref{genlower} and \ref{sec:euc} for proving lower bounds. Section
\ref{genlower} is devoted to lower bounds. In Section
\ref{sechightr} we analyze trees that have depth $h \ge \log n$. In
Section \ref{sublowtr} we turn to the complementary range $h < \log
n$. In Section \ref{sec:euc} we use the lower bounds for LLTs proven
in Section \ref{genlower} to derive our lower bounds on the tradeoff
between the hop-diameter and weight for Euclidean spanners. Our
upper bounds for LLTs are presented in Section \ref{sec:ub}. In
Section \ref{sec:sllt} these upper bounds are employed for
constructing SLLTs. Some basic properties of the binomial
coefficients that we use in our analysis appear in Appendix
\ref{app}.

\section {Preliminaries}
\label{sec:prel} For a positive integer $n$, an \emph{$n$-point
metric space} $M = (V,dist_M)$ can be viewed as the complete graph
$G=G(M) = (V,{V \choose 2},dist_M)$ in which for every pair of
vertices $u,w \in V$, the weight of the edge $e = (u,w)$ between $u$
and $w$ in $G$ is defined by $\omega(u,w) = dist_M(u,w)$. The
distance function $dist_M$ is required to be non-negative, equal to
zero when $u=w$, and to satisfy the triangle inequality
($dist_M(u,w) \le dist_M(u,v) + dist_M(v,w)$, for every triple
$u,w,v \in V$ of vertices). A graph $G'$ is called a \emph{spanning
subgraph} (respectively, \emph{spanning tree}; \emph{minimum
spanning tree}) of $M$ if it is a spanning subgraph (resp., spanning
tree; minimum spanning tree) of $G(M)$.

For a weighted graph $G = (V,E,\omega)$, and a path $P$ in $G$, its
\emph{(weighted) length} is defined as the sum of the weights of
edges along $P$, and its \emph{unweighted length} (or
\emph{hop-length}) is the number $|P|$ of edges (or \emph{hops}) in
$P$.
For a pair of vertices $u,w \in V$, the \emph{weighted}
(respectively, \emph{unweighted) distance in $G$ between $u$ and
$w$}, denoted $dist_G(u,w)$
(resp., $d_G(u,w)$), is the smallest weighted (resp.,
unweighted) length of a path connecting between $u$ and $w$ in $G$.
The \emph{weighted} (respectively, \emph{unweighted} or \emph{hop-})
\emph{diameter} of $G$ is the maximum weighted (resp., unweighted)
distance between a pair of vertices in $V$.

Whenever $n$ can be understood from the context, we write
$\vartheta$ as a shortcut for $\vartheta_n$. We will use the notion
{\em $\vartheta$-tree} as an abbreviation for a ``rooted spanning
tree of $\vartheta$".
We say that an edge $(v_i,v_j)$ connecting a parent vertex $v_i$
with a child vertex $v_j$ in a $\vartheta$-tree is a \emph{right}
(respectively, \emph{left}) \emph{edge} if $i > j$ (resp., $i < j$).
In this case $v_j$ is called a \emph{right} (resp., \emph{left})
\emph{child} of $v_i$. An edge $(v_i,v_j)$ is said to \emph{cover} a
vertex $v_{\ell}$ if $i < \ell < j$. For a $\vartheta$-tree $T$, the
number of edges $e \in E(T)$ that cover a vertex $v$ of $\vartheta$
is called the \emph{covering of $v$ by $T$} and it is denoted
$\chi(v) = \chi_{T}(v)$. The \emph{covering of the tree $T$},
$\chi(T)$, is the maximum covering of a vertex $v$ in $\vartheta$ by
$T$, i.e.,
$$\chi(T) = \max \{\chi_T(v) ~\vert~ v \in V(\vartheta)\}.$$
For a pair of positive integers $n$ and $h$, $1 \le h \le n-1$,
denote by $\chi(h)$ (respectively, $W(n,h)$) the minimum (vertex)
covering (resp., weight) taken over all $\vartheta_n$-trees of depth
$h$.

As was shown in \cite{AWY05}, the notions of covering and lightness
are closely related.

Finally, for a pair of non-negative integers $k,n$, $k \le n$, we
denote the sets $\{k,k+1,\ldots,n\}$ and $\{1,2,\ldots,n\}$
by $[k,n]$ and $[n]$, respectively.

\section {Monotonicity of Weight and Covering}
\label{secmon}

In this section we restrict our attention to $\vartheta$-trees and
show that both the minimum covering and the minimum weight do not
increase as the tree depth grows. This property is very useful for
proving lower bounds.

Fix a positive integer $n$. In what follows we write $\chi(h)$
(respectively, $W(h)$) as a shortcut for $\chi(n,h)$
(resp.,$W(n,h)$).

\begin {lemma} \label{moncov}
The sequence $(\chi(1),\chi(2),\ldots,\chi(n-1))$ is monotone
non-increasing.
\end {lemma}
\proof
Observe that $\chi(n-1)=0$, and for $h \ge 1$, $\chi(h)$ is
non-negative. Consequently, we henceforth restrict our attention to
the subsequence $(\chi(1), \chi(2),\ldots,\chi(n-2))$.

\begin{figure*}[htp]
\begin{center}
\begin{minipage}{\textwidth} 
\begin{center}
\setlength{\epsfxsize}{6.5in}
\epsfbox{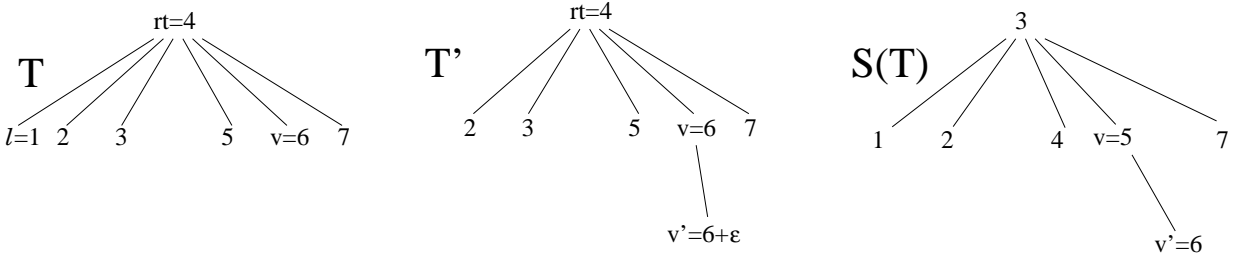}
\end{center}
\end{minipage}
\caption[]{ \label{figcover} \sf The $\vartheta_7$-tree $T$ rooted
at $rt=4$ is depicted on the left. This tree has depth 1 and
covering 2. The tree $T'$ on the vertices
$\{2,3,4,5,6,6+\epsilon,7\}$ rooted at $rt=4$ is depicted in the
middle. It has depth 2 and covering 2. This tree is obtained from
$T$ by removing the vertex $\ell=1$ along with the edge $(rt,\ell)$,
and adding the vertex $v' = (6+\epsilon)$ along with the edge
$(v,v')$. The $\vartheta_7$-tree $S(T)$ rooted at $3$ is depicted on
the right. It has depth 2 and covering 2. This tree is obtained from
$T'$ by relocating the points $2,3,4,5,6,6+\epsilon,7$ to the points
$1,2,\ldots,7$, respectively. }
\end{center}
\end{figure*}

Let $T$ be a $\vartheta$-tree that has depth $h$, $1 \le h \le n-2$,
and covering $\chi = \chi(h)$. (In other words, the tree $T$ has the
minimum covering among all trees of depth equal to $h$.) We denote
its root by $rt$. We construct a tree $S(T)$ that has depth $h+1$
and covering at most $\chi$. Consider a vertex $v$ at distance $h$
from $rt$, and the path $P=(rt=v_0,v_1,\ldots,v_{h-1},v)$ between
them in $T$.
\begin{enumerate}
\item Since $h \le n-2$, there exists at least one leaf $\ell$ in $T$ which
is not in $P$. Remove $\ell$ along with the edge connecting it to
its parent in $T$.
\item Let $\epsilon$, $0 < \epsilon < 1$, be a small real value.
Assume that $v < v_{h-1}$ (respectively, $v > v_{h-1}$). Insert a
new vertex $v'$, $v'=v-\epsilon$ (resp., $v'=v+\epsilon$) to be the
left (resp., right) child of $v$.
\end {enumerate}
Denote the resulting tree by $T'$. Note that $|V(T')| = |V(T)
\setminus \{\ell\} \cup \{v'\}| = n$. Clearly, the first step
neither changes the depth $h$ nor increases the covering $\chi$.
Since the distance from $rt$ to the farthest vertex $v$ in $T$ is
$h$, adding $v'$ as the left (resp., right) child of $v$ in the
second step increases the depth of the tree by exactly one. Note
that since $\epsilon < 1$, the new edge $(v,v')$ does not cover any
vertex in $T'$. Hence the covering of any vertex $v \in V(T')
\setminus \{v'\}$ in $T'$ is no greater than its covering in $T$. To
conclude that the covering of $T'$ is at most $\chi$, we show that
the covering of the new vertex $v'$ in $T'$ is no greater than
$\chi$. In fact, we argue that any edge that covers $v'$ also covers
$v$ in $T'$, which provides the required result. To see this, note
that for an edge $e$ that covers $v'$ not to cover $v$, it must hold
that $e$ is incident to $v$. However, since $v$ is a leaf in $T$,
the only edges which are incident to $v$ in $T'$ are $(v_{h-1},v)$
and the new edge $(v,v')$, both of which do not cover $v'$, and we
are done. (See Figure \ref{figcover} for an illustration.)

Observe that $T'$ does not span $\vartheta$. Let $v_1 < v_2 < \ldots
< v_n$ be the sequence of vertices of $T'$ in an increasing order.
To transform $T'$ into a spanning tree $S(T)$ of $\vartheta$, for
each index $i$, $1 \le i \le n$, relocate $v_i$ to the point $i$.

Let $T^*$ be a spanning tree that has depth $h+1$ and minimum
covering $\chi(h+1)$. By definition, $\chi(h+1)$ is no greater than
the covering of $S(T)$, which is at most $\chi(h)$. Hence $\chi(h+1)
\le \chi(h)$, and we are done. \QED

The following statement is analogous to Lemma \ref{moncov}. Its
proof is very similar (and, in fact, simpler) than that of Lemma
\ref{moncov} and is therefore omitted.
\begin {lemma} \label{monwt}
The sequence $(W(1),W(2),\ldots,W(n-1))$ is monotone non-increasing.
\end {lemma}
We remark that the monotonicity properties derived in this section
apply to any 1-dimensional Euclidean space (rather than just to
$\vartheta$).

\section {Lower Bounds}
\label{genlower}

In this section we devise lower bounds for lightness of
$\vartheta$-trees for the entire range of parameters. In Section
\ref{sechightr} we analyze trees of depth $h \ge \log n$ (henceforth
\emph{high} trees), and in Section \ref{sublowtr} we study trees
with depth in the complementary range $h < \log n$ (henceforth
\emph{low} trees).

\subsection {High Trees} \label{sechightr}

In this section we devise lower bounds for high $\vartheta$-trees.
In Sections \ref{secminla} and \ref{seccostfa} we restrict our
attention to binary trees, reduce this restricted variant of the
problem to a certain question concerning the minimum linear
arrangement problem, and resolve the latter question. In Section
\ref{seclbarb} we show that  the lower bound for binary trees
extends to general high trees, and, in fact, to general spanning
subgraphs.

\subsubsection {The Minimum Linear Arrangement Problem}
\label{secminla} In this section we describe a relationship between
the problem of constructing LLTs and the \emph{minimum linear
arrangement} (henceforth, \emph{MINLA}) problem \cite{ENRS95,RR98}.

The MINLA problem is defined as follows. Given an undirected graph
$G=(V,E)$, we would like to find a permutation (called also a
\emph{linear arrangement}) of the nodes $\sigma: V \rightarrow
\{1,\ldots,n=|V|\}$ that minimizes the cost of the linear
arrangement $\sigma$,
$$LA(G,\sigma) = \sum_{(i,j) \in E} |\sigma(i)-\sigma(j)|.$$
The minimum linear arrangement of the graph $G$, denoted $MINLA(G)$,
is defined as the minimum cost of a linear arrangement, i.e.,
$$MINLA(G) = \min \{LA(G,\sigma) ~\vert~ \sigma \in S_n \},$$
where $S_n$ is the set of all permutations of $[n]$.

Let $G=(V,E)$ be an $n$-vertex graph. For a permutation $\sigma \in
S_n$, let $G_{\sigma}= ([n],E_{\sigma})$ denote the graph with
vertex set $[n]$ and edge set $E_{\sigma} = \{(\sigma(u),\sigma(w))
~\vert~ (u,w) \in E\}$, equipped with the weight function $w(i,j) =
|i-j|$ for every $i \ne j$, $i,j \in [n]$. ($G_{\sigma}$ is an
isomorphic copy of $G$.) Observe that $LA(G,\sigma) =
\omega(G_{\sigma}) = \sum_{e \in E_{\sigma}} \omega(e)$. Also, let
$G^* = ([n],E^*)$ denote the graph $G_{\sigma^*}$ for the
\emph{optimal} permutation $\sigma^* = \sigma(G)$, that is, for
$\sigma^*$ such that $\omega(G_{\sigma^*}) = \min \{
\omega(G_{\sigma}) \vert \sigma \in S_n\}$. It follows that
$MINLA(G)$ is equal to $\omega(G_{\sigma^*})$. Moreover, for a
family $\mathcal F$ of $n$-vertex graphs, the minimum weight
$\omega(\mathcal F) = \{\omega(G_{\sigma^*}) \vert G \in \mathcal F
\}$ is precisely equal to the minimum value $MINLA(\mathcal F) =
\{MINLA(G) \vert G \in \mathcal F\}$ of the MINLA problem on one of
the graphs of the family $\mathcal F$. Next, we study the family
$B_n(h)$ of binary $\vartheta$-trees of depth no greater than $h$,
and show a lower bound on the value $Bin(n,h)$ of the minimum weight
of a $\vartheta$-tree from $B_n(h)$. Observe that
\begin{equation} \label{eq:binnh} Bin(n,h) ~=~ MINLA(B_n(h)).
\end {equation} Hence, it is sufficient to provide a lower bound for the
value $MINLA(B_n(h))$ of the MINLA for graphs of this family.

In a seminal work on spreading metrics, Even et al. \cite{ENRS95}
studied the following linear program relaxation $LP1$ for the MINLA
problem. The variables of this linear program $\{\ell(e) ~\vert~ e
\in E\}$ can be viewed as edge lengths. For a pair of vertices $u$
and $v$, $dist_\ell(u,v)$ stands for the distance between $u$ and
$v$ in the
graph $G$ equipped with length function $\ell(\cdot)$ on its edges. \\
$$\begin{array}{ll} {\displaystyle LP1: ~~\min ~~~\sum_{e \in E} \ell(e)} \cr\cr
{\displaystyle ~~~~~~~~~~~s.t.\ ~~~~\forall U \subseteq V, \forall v \in U ~:~
\sum_{u \in U} dist_\ell(v,u) \geq \frac{1}{4} \cdot (|U|^2 -1)} \cr\cr
~~~~~~~~~~~~~~~~~~~~{\displaystyle \forall e \in E ~:~ \ell(e) \ge 0.}
\end{array}$$

It is well-known that the optimal solution of this linear program is
a lower bound on $MINLA(G)$ \cite{ENRS95,RR98}.

As was already mentioned, we are only interested in the MINLA
problem for binary trees. Next, we present a variant $LP2$ of the
linear program $LP1$ which involves only a small subset of
constraints that are used in $LP1$. Consequently, the optimal
solution of $LP2$ is a lower bound on the optimal solution of $LP1$.
Consider a rooted tree $T=(T,rt)$. For a vertex $v$ in $T$, let
$U_v$ be the vertex set of the subtree of $T$ rooted at $v$. While
in $LP1$ there is a constraint for each pair $(U,v)$, $U \subseteq
V$, $v \in U$, there are only the constraints that correspond to
pairs $(U_v,v)$ present in $LP2$.
$$\begin{array}{ll} {\displaystyle LP2: ~~\min ~~~\sum_{e \in E(T)}  \ell(e)} \cr\cr
{\displaystyle ~~~~~~~~~~~ s.t.\ ~~~~\forall v \in V ~:~ \sum_{u \in
U_v} dist_\ell(u,v) \geq \frac{1}{4} \cdot (|U_v|^2 -1)} \cr\cr
~~~~~~~~~~~~~~~~~~~~~{\displaystyle \forall e \in E ~:~ \ell(e) \ge
0.}
\end{array}$$
\\

We will henceforth use the shortcut $dist(u,v)$ for
$dist_\ell(u,v)$.
For a vertex $v$ in $T$, let
$TotalDist(v) = \sum_{u \in U_v} dist_{\ell}(v,u)$, $Ineq(v)$
 be the inequality $TotalDist(v) \ge
\frac{1}{4} (|U_v|^2 -1)$, and $Eq(v)$ be the equation
$TotalDist(v) = \frac{1}{4} (|U_v|^2 -1)$.

 Next, we restrict our attention to binary trees.
 The
next lemma shows that if all inequalities $Ineq(v)$ are replaced by
equations
 $Eq(v)$, the value of the linear program $LP2$ does not change.
\begin {lemma} \label{equ}
For a binary $\vartheta$-tree $T$, in any optimal solution to $LP2$
all inequalities $\{Ineq(v) ~\vert~ v \in V\}$ hold as equalities.
\end {lemma}
{
\proof
First, observe that for a leaf $z$ in
$T$,
$$ TotalDist(z) ~=~ \sum_{u \in U_z} dist(u,z) ~=~ 0,$$ implying that
$Ineq(z)$ holds as equality.
\\ Let $n$ denote the number of vertices of $T$. Order the $(n-1)$ edges $e_1,e_2,\ldots,e_{n-1}$
arbitrarily, and consider the subset $\mathcal C$ of all value
assignments $\psi$ to the variables
$\ell(e_1),\ell(e_2),\ldots,\ell(e_{n-1})$, that constitute an
optimal solution to the linear program $LP2$, and such that there
exists a vertex $v \in V(T)$ for which $Ineq(v)$ holds as a strict
inequality under $\psi$. Suppose for contradiction that $\mathcal C
\ne \phi$. For an assignment $\psi \in \mathcal C$, let the
\emph{level} of $\psi$, denoted $L(\psi)$, be the minimum level of a
vertex $v$ in $T$ for which $Ineq(v)$ holds as a strict inequality.
Consider a optimal solution $\psi^* \in \mathcal C$ of minimum
level, that is,
$$L(\psi^*) ~=~ \min\{L(\psi) ~\vert~ \psi \in \mathcal C\}.$$ Let $v$ be an
inner vertex of level
$L(\psi^*)$ for which $Ineq(v)$ holds as a strict inequality
under the assignment $\psi^*$. By definition,
$$TotalDist(v) ~>~ \frac{1}{4} (|U_v|^2 -1).$$

It is convenient to imagine that the vertices of $T_v$ are
colored in two colors as follows. The root vertex $v$ of $T_v$ is
colored white. All leaves are colored black. An inner vertex $u
\in U_v \setminus \{v\}$ is colored white, if the following three
conditions hold.
\begin {itemize}
\item Its parent $\pi(u)$ in $T_v$ is colored white.
\item $Ineq(u)$ holds as a strict inequality.
\item All edges $e$ connecting
$u$ to its children satisfy $\ell(e) = 0$ under $\psi^*$.
\end {itemize}
Otherwise, $u$ is colored black.
\\{\bf Remark:} Observe that for a white vertex $u \in U_v \setminus
\{v\}$, all vertices of the path $P_{v,\pi(u)}$ connecting $v$ with $\pi(u)$ are colored
white.

\begin {claim} \label{nested}
At least one vertex of depth 1 in $T_v$ is colored white.
\end {claim}
\proof
Suppose for contradiction that all white vertices
in $T_v$ have depth at least 2.
Let $w$ be a white vertex of minimum depth $d$, $d \ge 2$.
 Since $w$ is colored white,
all vertices in the path $P_{v,w}$ connecting $v$ with $w$ in $T_v$ are
colored white as well, implying that for each vertex $x$ along that
path, $Ineq(x)$ holds as a strict inequality, and all edges that
connect $x$ to its children have weight zero. 

Denote the left (respectively, right) child of $w$ by $w_L$ (resp.,
$w_R$). Since the weight of the edges that connect $w$ to its
children have weight zero, it holds that
$$TotalDist(w) = TotalDist(w_L) + TotalDist(w_R).$$ Since
$Ineq(w)$ holds as a strict inequality,
$$\begin{array}{ll}
{\displaystyle TotalDist(w) ~>~ \frac{1}{4} \cdot (|U_w|^2-1) ~=~
 \frac{1}{4} \cdot ((|U_{w_L}|+|U_{w_R}|+1)^2-1)} \cr\cr {\displaystyle ~~~~~~~~~~~~~~~~~~~>~
\frac{1}{4} \cdot (|U_{w_L}|^2 -1)+ \frac{1}{4} \cdot (|U_{w_R}|^2-1).}
\end{array}$$ Thus, at
least one among the two inequalities $Ineq(w_L)$ and $Ineq(v_R)$
holds as a strict one. We assume without
loss of generality that $Ineq(w_L)$ holds as a strict inequality.

To complete the proof we need the following claim.
\begin {claim}
All edges that connect $w_L$
to its children have value zero under $\psi^*$.
\end {claim}
\proof Suppose for contradiction that there is a child $y$ of $w_L$
such that the length of the edge $e=(w_L,y)$ under the assignment
$\psi^*$ is some $\delta
>0$. Consider the path $P_{v,w} = (v=v_0,v_1,\ldots,v_j=w)$, $j
\ge 0$, connecting the vertices $v$ and $w$. The analysis splits
into two cases depending on whether $v$ is the root $rt$ of $T$ or
not. First, suppose that $v=rt$. Observe that for every index
$i$, $i \in [0,j]$,
$$f_i(\delta) ~=~ f_i(\ell(e)) ~=~ TotalDist(v_i) - \frac{1}{4} \cdot
(|U_{v_i}|^2-1)$$ is a continuous function of the variable
$\ell(e)$. Since for every $i \in [0,j]$, $f_i(\delta) > 0$, we can
slightly decrease the value of $\ell(e)$ and set it to some
$\delta'$, $0 < \delta' < \delta$, so that all $f_i(\delta')$ are
still non-negative. However, this change in the value of $\ell(e)$
results in a new feasible assignment $\psi'$ of values to the
variables $\{\ell(e) \vert e \in E(T)\}$. Moreover, obviously
$\sum_{e \in E(T)} \ell(e)$ of the objective function of $LP2$ is
smaller under $\psi'$ than under $\psi^*$. This is a contradiction
to the assumption that $\psi^*$ is a optimal solution for $LP2$.

The case that $v \ne rt$ is handled similarly. In this case the
difference $\epsilon = \delta - \delta'$ is added to the value of
$\ell(e')$, for the edge $e' = (v,\pi(v))$ connecting $v$ to its
parent in $T$. It is easy to verify that the resulting assignment
$\tilde{\psi}$ is feasible, and that the value of the objective
function $\sum_{e \in E(T)} \ell(e)$ is the same under $\psi^*$ and
$\tilde{\psi}$. Also, since for every $i \in [0,j]$, $f_i(\ell(e)) =
f_i(\delta') > 0$ under $\tilde{\psi}$, it follows that the
inequalities $Ineq(v_i)$ hold as strict inequalities for all $i \in
[0,j]$, and thus both assignments $\tilde{\psi}$ and $\psi^*$ belong
to the set $\mathcal C$. However, $Ineq(\pi(v))$ holds as a strict
inequality under $\tilde{\psi}$ as well, and thus $L(\tilde{\psi}) <
L(\psi^*)$. This is a contradiction to the assumption that $\psi^*$
has the minimum level in $\mathcal C$. Hence under $\psi^*$, all
edges that connect $w_L$ to its children have value zero. \QED

Recall that $Ineq(w_L)$ holds as a
strict inequality, and $w = \pi(w_L)$ is a white vertex.
Consequently, $w_L$ should be colored white as well. However, its
depth is smaller than the minimum depth of a white vertex in
$T_v$, contradiction.
This completes the proof of Claim \ref{nested}.
\QED

Consider a white vertex $x$ in $T_v$ of depth 1. The edges
connecting $x$ to its children are assigned value zero, implying
that $TotalDist(x) = 0$. However, since $x$ is colored white, the
inequality $Ineq(x)$ holds as a strict inequality, i.e.,
$$TotalDist(x) ~>~ \frac{1}{4} \cdot (|U_x|^2-1) ~>~ 0.$$
This is a contradiction to the assumption that $\mathcal C$ is not
empty, proving Lemma \ref{equ}.
\QED
}

Consider a subtree $T_v$ rooted at an inner vertex $v$. Without
loss of generality, $v$ has a left child $v_L$, and possibly a
right child $v_R$, each being the root of the corresponding
subtrees $T_{v_L}$ and $T_{v_R}$, respectively. Let $U_L =
U_{v_L}$, $U_R = U_{v_R}$, $n_L = |U_{L}|$, $n_R = |U_{R}|$, $e_L
= (v,v_L)$, $e_R = (v,v_R)$, $x_L = \ell(e_L)$, and $x_R =
\ell(e_R)$. If $v$ has only a left child, then we write $v_R =
T_{v_R} = U_{v_R} = NULL$, and $n_R= x_R =0$. Also, without loss
of generality assume that $n_L \ge n_R$.

The next lemma provides a lower bound on the sum $x_L + x_R$ of
values assigned by a minimal optimal solution for $LP2$ to the edges $e_L$ and $e_R$.
\begin {lemma} \label{costing}
For an optimal solution for $LP2$,
 $$x_L + x_R ~>~ \frac{1}{2} \cdot
(\min\{n_L,n_R\} +1) ~=~ \frac{1}{2} \cdot (n_R +1).$$

\end {lemma}
{
\proof
It is easy to verify that \begin {equation} \label {disteq}
\sum_{u \in U_v} dist(v,u) ~=~ x_L \cdot n_L + \sum_{u \in U_{L}}
dist(v_L,u) + x_R \cdot n_R + \sum_{u \in U_{R}} dist(v_R,u). \end
{equation} By Lemma \ref{equ}, both inequalities $Ineq(v)$ and
$Ineq(v_L)$ hold as equalities, i.e,
\begin {eqnarray} \label{fur}
\sum_{u \in U_v} dist(v,u) &=& \frac{1}{4} \cdot (|U_v|^2-1) ~=~
\frac{1}{4} \cdot ((n_L+n_R+1)^2-1). \\
\label{suc}
\sum_{u \in U_{L}} dist(v_L,u) &=& \frac{1}{4} \cdot ({n_L}^2-1).
\end {eqnarray}

The analysis splits into two cases.
\\\emph{Case 1: $v$ has two children.}
By Lemma \ref{equ}, the inequality $Ineq(v_R)$ holds as equality
as well, i.e.,
\begin {equation} \label{thur}
\sum_{u \in U_{R}} dist(v_R,u) ~=~  \frac{1}{4} \cdot
({n_R}^2-1). \end {equation}


 Plugging the equations (\ref{fur}), (\ref{suc}), and (\ref{thur}) in equation
(\ref{disteq}) implies
\begin {equation} \label{xl}
x_L \cdot n_L + x_R \cdot n_R = \frac{1}{2} \cdot (n_L \cdot n_R +
(n_L + n_R) + 1).
\end {equation}

Since $n_L \ge n_R$, and $n_L >0$, it follows that
$$
x_L  + x_R \cdot \frac{n_R}{n_L}
 ~>~ \frac{1}{2} \cdot (n_R +1),$$
and so,
$$x_L + x_R ~\ge~ x_L  + x_R \cdot \frac{n_R}{n_L} ~>~
\frac{1}{2} \cdot (n_R + 1) ~=~ \frac{1}{2} \cdot (\min\{n_L,n_R\} +
1).$$
\\\emph{Case 2: $v$ has only a left child.}
Then $n_R = x_R = 0$, and \begin {equation} \label{castwo} \sum_{u
\in U_v} dist(v,u) ~=~ x_L \cdot n_L + \sum_{u \in U_{L}}
dist(v_L,u).
\end {equation}
Plugging equations (\ref{fur}) and (\ref{suc}) in equation
(\ref{castwo}), we obtain
$$x_L \cdot n_L ~=~ \frac{1}{4} \cdot (2\cdot n_L+1).$$ Hence
$$x_L+x_R ~=~ x_L
~>~ \frac{1}{2} ~=~ \frac{1}{2} \cdot (\min\{n_L,n_R\} + 1).$$
\QED
}

\subsubsection {The Cost Function} \label{seccostfa}

In this section
we define and analyze a cost function on binary
$\vartheta$-trees. 
We will show that in order to provide a lower
bound for $MINLA(B_n(h))$, it is sufficient to provide a lower bound for the
minimum value of this cost function  on a tree from $B_n(h)$.

 Consider a binary $\vartheta$-tree $(T,rt)$ in which for every
inner vertex $v$ that has two children, one of those children is
designated as the \emph{left} child $v.left$ and the other as the
\emph{right} one $v.right$. If $v$ has just one child then this
child is designated as the left one. Also, for an inner vertex $u$,
let $|u|$ denote the number of vertices in the subtree of $T$ rooted
at $u$. Let $I=I(T)$ denote the set of inner vertices of $T$.
 By Lemma \ref{costing}, for any
optimal assignment $\psi$ for the values $\{\ell(e) ~\vert~ e \in
E(T) \}$ of the linear program $LP2$,
\begin {equation} \label{ocost}
\sum_{e \in E(T)} \ell(e) ~=~ \sum_{v \in I(T)}
\left(\ell(v,v.left)+\ell(v,v.right)\right) ~\ge~ \frac{1}{2}
\cdot \sum_{v \in I(T)} (\min\{|v.left|,|v.right|\}+1).
\end {equation}
We call the right-hand side expression the \emph{cost} of the tree
$T$, and denote it $Cost(T)$. Let $MINCOST(B_n(h))$ denote
$\min\{Cost(T) \vert T \in B_n(h)\}$. It follows that $MINLA(B_n(h))
\ge MINCOST(B_n(h))$, and in the sequel we provide a lower bound for
$MINCOST(B_n(h))$. Note that by (\ref{eq:binnh}), this lower bound
will apply to $Bin(n,h)$ as well.

The subtree
rooted at the left (respectively, right) child of $T$ is called
the \emph{left subtree} (resp., \emph{right subtree}) of $T$. We
will use the notation $T.left$ and $rt.left$ (respectively,
$T.right$ and $rt.right$) interchangeably to denote the left
(resp., right) subtree of $T$. Also, let $|T|$ denote the
\emph{size} of the tree $T$, that is, the number of vertices in
$T$.\\Consider the following cost function on binary trees,
$$Cost'(T) ~=~ Cost'(T.left) + Cost'(T.right) + \min\{|T.left|,|T.right|\}.$$
It is easy to verify that $Cost(T)$ can be equivalently expressed as
$$
Cost'(T) ~=~ \sum_{v \in I(T)} \min\{|v.left|,|v.right|\}.
$$
 Since for
any binary tree $T$, $2 \cdot Cost(T) \ge Cost'(T)$, we will
henceforth focus on proving a lower bound for $Cost'(T)$, and use
the notion ``cost" to refer to the function $Cost'$. Fix a pair of
positive integers $n$ and $h$, $n-1 \ge h$. A rooted binary tree on
$n$ vertices that has depth at most $h$ will be called an
\emph{(n,h)-tree}. Let $R(n,h)$ denote the minimum cost taken over
all $(n,h)$-trees. It follows that
\begin{equation} \label{eq:binrnh}
Bin(n,h) ~\ge~ \frac{1}{2} \cdot R(n,h).
\end {equation}

This section is devoted to proving
the following theorem that establishes
lower bounds on $R(n,h)$, for all $h \ge \log n$.

\begin {theorem} \label {majort}
\begin {enumerate}
\item If $\log n \le h \le 2 \lfloor  \log n \rfloor$, then $R(n,h)  \ge \frac{2}{3} \cdot n \cdot \lfloor \frac{1}{8}\log n \rfloor.$
\item If  $~2 \lfloor  \log n \rfloor < h \le n-1$, let $f(h)$ be the minimum integer such that ${h+1 \choose f(h)} > \frac{2}{3} \cdot n$.
    Then $R(n,h) > \frac{2}{3} \cdot n \cdot (f(h)-2)$.
\end {enumerate}
\end {theorem}
{\bf Remark 1:} Note that for $h > 2 \lfloor  \log n \rfloor$, ${h+1
\choose  \log n } > \frac{2}{3} \cdot n$, and thus $f(h)$ is
well-defined in this range. \\
{\bf Remark 2:} By (\ref{eq:binrnh}), the lower bounds of
Theorem \ref{majort} apply (up to a factor of 2) to $Bin(n,h)$ as
well.

\begin{figure*}[htp]
\begin{center}
\begin{minipage}{\textwidth} 
\begin{center}
\setlength{\epsfxsize}{4.5in} \epsfbox{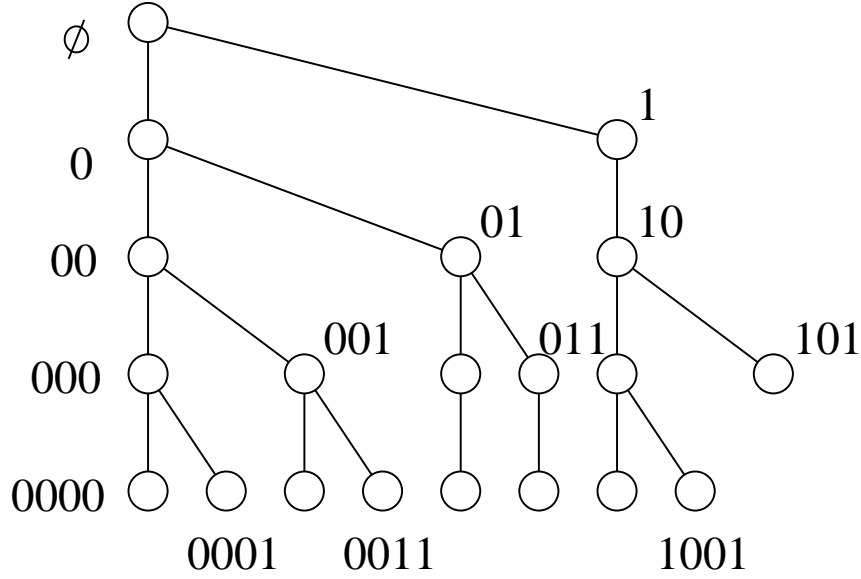}
\end{center}
\end{minipage}
\caption[]{ \label{fig:bin} \sf A cost-optimal binary tree for
$n=20$ and $h=4$. }
\end{center}
\end{figure*}

Let $n$ and $h$ be non-negative integers. Given a binary tree $T$,
we restructure it \emph{without changing its cost} and depth, so
that for each vertex $v$ in $T$, the size of its right subtree
$v.right$ would not exceed the size of its left subtree $v.left$.
Specifically, if in the original tree $T$ it holds that $|v.left|
\ge |v.right|$, then no adjustment occurs in $v$. However, if
$|v.left| < |v.right|$, then the restructuring process exchanges
between the left and right subtrees of $v$. We refer to this
restructuring procedure as the \emph{right-adjustment} of $T$, and
denote the resulting binary tree by $\tilde{T}$. (See Figure
\ref{fig:bin} for an illustration.) Since $T$ and its
\emph{right-adjusted} tree $\tilde{T}$ have the same cost, we
henceforth restrict our attention to right-adjusted trees. By
definition, in a right-adjusted tree $\tilde T$, for any $v \in
V(\tilde{T})$, it holds that $|v.right| \le |v.left|$, and
consequently, $Cost(\tilde{T}) = {\displaystyle \sum_{v \in
V(\tilde{T})} |v.right|}$.

A set of $n$ binary words with at most $h$ bits each will be called
an \emph{(n,h)-vocabulary}. Next, we define an injection
$\mathcal{S}$ from
the set of $(n,h)$-trees to the set of $(n,h)$-vocabularies.\\
For a vertex $v$ in a binary tree $T$, denote by $P_v =
(rt=v_0,v_1,\ldots,v_k=v)$ the path from $rt$ to $v$ in $T$, and
define $B_v = b_1b_2\ldots b_k$ to be its corresponding binary word,
where for $i \in \{1,2,\ldots,k\}$, $b_i = 0$ if $v_{i}$ is the left
child of $v_{i-1}$, and $b_i=1$ otherwise. Given an $(n,h)$-tree
$T$, let $\mathcal S(T)$ be the $(n,h)$-vocabulary that consist of
the $|T|$ binary words that correspond to the set of all
root-to-vertex paths in $T$, namely, $\mathcal S(T) = \{B_v ~\vert~
v \in V(T)\}$. (See Figure \ref{fig:bin} for an illustration.)
\\For a binary
word $W$, denote its \emph{Hamming weight} (the number of 1's in it)
by $H(W)$. For a set $S$ of binary words, define its total Hamming
weight, henceforth \emph{Hamming cost}, $HCost(S)$, as the sum of
Hamming weights of all words in $S$, namely, $HCost(S)=
{\displaystyle \sum_{B \in S} H(B)}$. Finally, denote the minimum
Hamming cost of a set of $n$ distinct binary words with at most $h$
bits each by $H(n,h)$. Observe that the function $H(n,h)$ is
monotone non-increasing with $h$.

In the next lemma we show that it is sufficient to prove the desired
lower bound for $H(n,h)$.
\begin {lemma} \label{mono}
For all positive integers $n$ and $h$, $n-1 \ge h$, $H(n,h) \le
R(n,h)$.
\end {lemma}
{
\proof
It is easy to verify by a double counting that in a right-adjusted
tree $\tilde{T}$,
$$\sum_{v \in V(\tilde{T})} |v.right|
~=~  \sum_{v \in V(\tilde{T})} | \{u \in V(P_v) :  v \in
u.right\}|.$$ Consequently,
$$
\begin{array}{ll}
Cost(\tilde{T}) ~=~ {\displaystyle \sum_{v \in V(\tilde{T})}}
|v.right| ~=~ {\displaystyle \sum_{v \in V(\tilde{T})}} | \{u \in
V(P_v) ~:~  v \in u.right\}| \cr\cr~~~~~~~~~~~ ~=~ {\displaystyle
\sum_{v \in V(\tilde{T})}} H(B_v)
 ~=~ HCost(\mathcal{S}(\tilde{T})).
\end{array}
$$

Let $T^*$ be a right-adjusted $(n,h)$-tree realizing $R(n,h)$, that
is, $Cost(T^*) = R(n,h)$. It follows that $$R(n,h) ~=~ Cost(T^*) ~=~
HCost(\mathcal{S}(T^*)) ~\ge~ H(n,h).$$
\QED
}

In what follows we establish lower bounds on $H(n,h)$.

Consider a set $\mathcal S^* = \mathcal S^*(n,h)$ realizing
$H(n,h)$, that is, a set that satisfies $HCost(\mathcal S^*) =
H(n,h)$. For a non-negative integer $i$, $i \le h$, let $\mathcal
S(h,i)$ be the set of all distinct binary words with at most $h$
bits each, so that each word of which contains precisely $i$ 1's. To
contain the minimum total number of 1's, the set $\mathcal S^*$
needs to contain all binary words with no 1's, all binary words that
contain just a single 1, etc. In other words, there exists an
integer  $r=r(h)$ for which
$$\bigcup_{i=0}^{r} \mathcal S(h,i) ~\subset~
 \mathcal S^* ~\subseteq~ \bigcup_{i=0}^{r+1} \mathcal S(h,i).$$
%
(See Figure \ref{fig:bin} for an illustration.)

By Fact \ref{fact23},
$$|\mathcal S(h,i)| ~=~ \sum_{k=i}^h {k
\choose i} ~=~ {h+1 \choose i+1}.$$
 Note that for a pair of distinct
indices $i$ and $j$, $0 \le i,j \le h$, the sets $\mathcal S(h,i)$
and $\mathcal S(h,j)$ are disjoint. Since $|\mathcal S^*| = n$, it
holds that
\begin {equation} \label{first}
\sum_{i=0}^{r} {h+1 \choose i+1} ~<~ n ~\le~ \sum_{i=0}^{r+1} {h+1
\choose i+1}.
\end {equation}
Recall that for every non-negative integer $i$, $i \le h$, each word
in $\mathcal S(h,i)$ contains precisely $i$ 1's, and thus
$$HCost(\mathcal S(h,i)) =  i \cdot {h+1 \choose i+1}.$$
Let
\begin {equation} \label{second}
N ~=~ n- \sum_{i=0}^{r} {h+1 \choose i+1} > 0
\end {equation}  be the number of words
with Hamming weight $r+1$ in $\mathcal S^*$. Hence
 \begin
{equation} \label{third}
\begin{array}{ll}
H(n,h) ~=~ HCost(\mathcal S^*)~=~ {\displaystyle \sum_{i=0}^{r}
HCost(\mathcal S(h,i))} + N \cdot (r+1) \cr\cr~~~~~~~~~~ ~=~
{\displaystyle \sum_{i=0}^{r} i \cdot {h+1 \choose i+1}} + N \cdot
(r+1).
\end{array}
\end {equation}
%
The next claim establishes a helpful relationship between the parameters
$h$, $r$, and $n$.
\begin {claim} \label{up1}
For $h \ge \lfloor 2 \log n \rfloor$, $r \le \left\lfloor
\frac{h+1}{4} \right\rfloor -1$.
\end {claim}
{
\proof
Suppose for contradiction that $r \ge \lfloor \frac{h+1}{4}
\rfloor$. Then $$r+1 \ge \left \lfloor \frac{\lfloor 2 \log n
\rfloor +1}{4} \right \rfloor +1.$$ Hence $$n \le {\lfloor 2 \log n
\rfloor+1 \choose \left \lfloor \frac{\lfloor 2 \log n \rfloor +1}{4}
\right\rfloor +1} \le {h+1 \choose r+1} < \sum_{i=0}^{r} {h+1
\choose i+1}.$$ However, by (\ref{first}) the right-hand
side is smaller than $n$, contradiction.
\QED
}


The next lemma establishes lower bounds on $H(n,h)$ for $h \ge
\lfloor 2 \log n \rfloor$. Since $H(n,h)$ is monotone non-increasing
with $h$, it follows that the lower bound for $h= 2 \lfloor \log n
\rfloor$ applies for all smaller values of $h$.
\begin {lemma} \label{major}
\begin {enumerate}
\item $H(n,2 \lfloor  \log n \rfloor) \ge \frac{2}{3} \cdot n \cdot \lfloor \frac{1}{8}\log n \rfloor.$
\item For  any $~2 \lfloor  \log n \rfloor < h \le n-1$, $H(n,h) > \frac{2}{3} \cdot n \cdot (f(h)-2)$.
\end {enumerate}
\end {lemma}
\proof
By Lemma \ref{lemm1},
$$\sum_{i=0}^{r} {h+1 \choose i+1} ~<~ \frac{3}{2} \cdot {h+1 \choose
  r+1}.$$
Since
$$\sum_{i=0}^{r} {h+1 \choose i+1} ~=~ n-N,$$
 it follows that
$${h+1 \choose r+1} ~>~ \frac{2}{3}\cdot (n-N),$$
 and so,
$${h+1 \choose
r+1} + N ~>~  \frac{2}{3}\cdot n.$$
 Hence, by
equation
(\ref{third}),
\begin {equation} \label{fifth}
H(n,h) ~=~ \sum_{i=0}^{r} i \cdot {h+1 \choose i+1} + N \cdot (r+1)
~>~ r\cdot {h+1 \choose r+1} +  N \cdot (r+1)~>~ \frac{2}{3} \cdot
n\cdot r.
\end {equation}
\\

The analysis splits into two cases.
\\ \emph{Case 1: $h= \lfloor 2 \log n \rfloor$.}
\\We will prove the first assertion of Lemma \ref{major}, that is,
$H(n,\lfloor 2 \log n \rfloor) \ge \frac{2}{3} \cdot n \cdot \lfloor
\frac{1}{8}\log n \rfloor$. We assume that $n \ge 256$, as otherwise
the right-hand side vanishes and the statement holds trivially.

The next claim shows that in this case $r$ is quite large.
\begin {claim}
\label{cl:rr}
 $r \ge \left\lfloor \frac{1}{8} \log n \right\rfloor$.
\end {claim}
{
\proof
Suppose for contradiction that $r \le \left\lfloor \frac{1}{8} \log
n \right\rfloor -1$. Observe that for $n \ge 3$,
$\left\lfloor \frac{1}{8} \log n \right\rfloor +1 ~\le~ \left\lfloor
\frac{\lfloor 2 \log n \rfloor + 1}{4} \right\rfloor,$ and thus Lemma \ref{lemm1}
is applicable. Hence,
\begin {equation} \label{equ1}
\sum_{i=0}^{\lfloor \frac{1}{8}\log n \rfloor+1} {\lfloor 2 \log n \rfloor+1 \choose i} ~<~
\frac{3}{2} \cdot {\lfloor 2 \log n \rfloor+1 \choose \lfloor
\frac{1}{8}\log n \rfloor +1}.
\end {equation}
 By (\ref{first}),
 $$
 \begin{array}{ll}
 {\displaystyle n ~\le~ \sum_{i=0}^{r+1} {h+1 \choose i+1} ~=~ \sum_{i=0}^{r+1} {\lfloor 2 \log n \rfloor+1 \choose i+1}} \cr\cr~~ {\displaystyle ~\le~  \sum_{i=0}^{\lfloor \frac{1}{8}\log n \rfloor} {\lfloor 2 \log n \rfloor+1 \choose i+1}
  ~=~ \sum_{i=0}^{\lfloor \frac{1}{8}\log n \rfloor+1} {\lfloor 2 \log n \rfloor+1 \choose i}
 -1}.
\end {array}$$
However, (\ref{equ1}) implies that the right-hand side is strictly
smaller than $\frac{3}{2} \cdot {\lfloor 2 \log n \rfloor+1 \choose \lfloor \frac{1}{8}\log n \rfloor +1} -1
~\le~ n,$
contradiction.
\QED
}

Consequently,
by (\ref{fifth}),
we have
$$H(n,\lfloor 2\log n \rfloor) > \frac{2}{3} \cdot n \cdot r \ge \frac{2}{3} \cdot n \cdot \lfloor \frac{1}{8}\log n \rfloor.$$

This completes the proof of the first assertion of Lemma
\ref{major}.
\\
To prove the second assertion we analyze the case $h
>  \lfloor 2 \log n \rfloor$.
$\\$ \emph{Case 2: $h >   \lfloor 2 \log n \rfloor.$} \\We start the
analysis of this case by showing that for $h$ in this range, the
upper bound on the value of $r$ established in Claim \ref{up1} can
be improved.
\begin {claim} \label{imp}
$r \le \left\lfloor \frac{h+1}{4} \right\rfloor -2$.
\end {claim}
{
\proof
It is easy to verify that $n \le {\lfloor 2 \log n \rfloor+2
\choose \left \lfloor \frac{\lfloor 2 \log n \rfloor +2}{4} \right
\rfloor}.$ Suppose for contradiction that $r \ge \left\lfloor
\frac{h+1}{4} \right\rfloor -1$.
 Then $r+1 \ge \left \lfloor \frac{\lfloor 2 \log n \rfloor +2}{4} \right \rfloor.$
Hence $$n \le {\lfloor 2 \log n \rfloor+2 \choose \left \lfloor
\frac{\lfloor 2 \log n \rfloor +2}{4} \right \rfloor} \le {h+1
\choose r+1} < \sum_{i=0}^{r} {h+1 \choose i+1}.$$ However, by
(\ref{first}), the right-hand side is smaller than $n$,
contradiction.
\QED
}

To complete the proof, we provide a lower bound on $r$.\\
Recall that $f(h)$ is defined to be the minimum integer that
satisfies ${h+1 \choose f(h)} > \frac{2}{3} \cdot n$.
\begin {claim} \label{fh}
${h+1 \choose r+2} > \frac{2}{3} \cdot n$.
\end {claim}
{
By Claim \ref{imp}, $r \le \lfloor \frac{h+1}{4} \rfloor -2$. Hence,
Lemma \ref{lemm1} implies that
$$\sum_{i=0}^{r+1} {h+1 \choose i+1} ~<~ \frac{3}{2} \cdot {h+1 \choose r+2}.$$
Consequently, by (\ref{first}),
$$n \le \sum_{i=0}^{r+1} {h+1 \choose i+1 } ~<~ \frac{3}{2} \cdot {h+1 \choose r+2}.$$
The assertion of the claim follows.
\QED
}

Claim \ref{fh} implies that $r \ge f(h)-2$.
By
 equation
(\ref{fifth}),
 it follows that
 $$H(n,h) > \frac{2}{3} \cdot n \cdot
r \ge \frac{2}{3} \cdot n \cdot (f(h)-2),$$
 implying the second
assertion of Lemma \ref{major}.
\QED

Lemmas \ref{mono} and \ref{major} imply Theorem \ref{majort}.
%
We are now ready to derive the desired lower bound for binary trees.

\begin {theorem} \label{asym}
For sufficiently large integers $n$ and $h$, $h \ge \log n$, the
minimum weight $Bin(n,h)$ of a  binary $\vartheta$-tree that has
depth at most $h$ is at least $\Omega(\Psi \cdot n)$, for some
$\Psi$ satisfying $h = \Omega(\Psi \cdot n^{1/\Psi})$.
\end {theorem}

{
\proof
 First, observe that $Bin(n,h) \ge \frac{1}{2}
\cdot R(n,h)$.
The analysis splits into
three cases, depending on the value of $h$.
\\\emph{Case 1: $\log n  \le h \le 2 \lfloor \log n \rfloor.$}
By the first assertion of Theorem \ref{majort}, $$Bin(n,h) ~\ge~
\frac{1}{2} \cdot R(n,h) ~=~ \Omega(\log n \cdot n).$$ Observe that
for $h \in \left[\log n,2 \lfloor \log n \rfloor\right]$, and $k =
\log n$, it holds that $h= \Omega\left(k \cdot n^{1/k}\right)$, and
we are done.
\\\emph{Case 2: $2 \lfloor \log n \rfloor \le h \le n^{1/4}$.} By the
second assertion of Theorem \ref{majort}, $$Bin(n,h) ~\ge~
\frac{1}{2} \cdot R(n,h) ~\ge~ \frac{1}{3} \cdot n \cdot (f(h)-2),$$
where $f(h)$ is the minimum integer that satisfies ${h+1 \choose
f(h)} > \frac{2}{3} \cdot n$. Observe that for a sufficiently large
$n$, and $h \in [2 \lfloor \log n \rfloor,n^{1/4}]$, we have that
$f(h) \ge 4$, and so, $f(h) -2 \ge \frac{1}{2} \cdot f(h)$. It
follows that $Bin(n,h) = \Omega(f(h) \cdot
n).$
 Also, it holds that $$\left(\frac{e \cdot (h+1)}{f(h)}\right)^{f(h)}
~>~ {h+1 \choose f(h)} ~>~ \frac{2}{3} \cdot n.$$ Consequently, $$h
~>~ \left(\frac{1}{e} \cdot f(h) \cdot \left(\frac{2}{3} \cdot n
\right)^{\frac{1}{f(h)}}\right) -1 ~=~ \Omega\left(f(h) \cdot
n^{\frac{1}{f(h)}}\right).$$
\emph{Case 3: $h > n^{1/4}$.}
In this range any constant $k \ge 4$ satisfies $h = \Omega(k \cdot n^{1/k})$.
Note that the weight of the minimum spanning tree
of $\vartheta$ is $n-1$, which implies that for a constant $k$,
$$Bin(n,h) ~\ge~ n-1 ~=~ \Omega(k \cdot n).$$
\QED
}




\subsubsection {Lower Bounds for General High Trees} \label{seclbarb}

 In this
section we show that our lower bound for binary trees implies an
analogous lower bound for general high trees. We do this in two
stages. First, we show that a lower bound for trees in which every
vertex has at most four children, henceforth \emph{$4$-ary trees},
will suffice. Second, we show that the lower bound for binary trees
implies the desired lower bound for $4$-ary trees.

For an inner node $v$ in a tree $T$, we denote its children by
$c_1(v),c_2(v),\ldots,c_{ch(v)}(v)$, where $ch(v)$ denotes the
number of its children in $T$. Suppose without loss of
generality that the children are ordered so that the sizes of
 their corresponding subtrees form a monotone non-increasing sequence, i.e., $|T_{c_1(v)}| \ge |T_{c_2(v)}| \ge \ldots \ge
|T_{c_{ch(v)}(v)}|$.
For a vertex $v$ in $T$, we define its
\emph{star subgraph} $S_v$ as the subgraph of $T$ connecting
$v$ to its children in $T$, namely, $S_v = (V_v,E_v)$, where $V_v
= \{v,c_1(v),\ldots,c_{ch(v)}(v)\}$ and $E_v = \{(v,c_i(v)) ~\vert~
i=1,2,\ldots,ch(v)\}$.
 For convenience, we denote $T_i = T_{c_i(rt)}$, for $i=1,2,\ldots,ch(v)$.

\begin{figure*}[htp]
\begin{center}
\begin{minipage}{\textwidth} 
\begin{center}
\setlength{\epsfxsize}{5.5in} \epsfbox{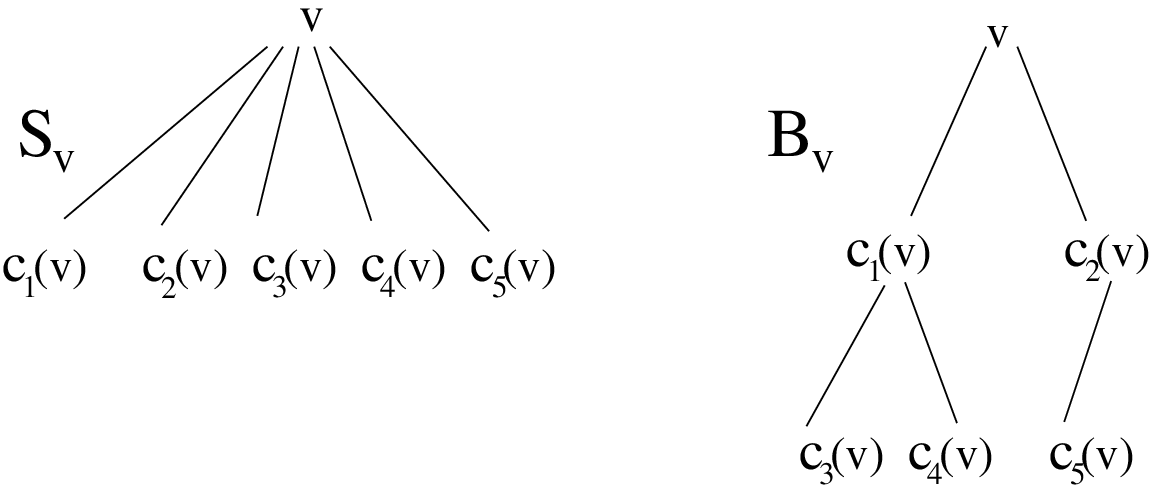}
\end{center}
\end{minipage}
\caption[]{ \label{full} \sf The tree $S_v$ on the left is the star
subgraph rooted at $v$, and having the five vertices
$c_1(v),c_2(v),\ldots,c_5(v)$ as its leaves. The full binary tree
$B_v$ on the right is obtained as a result of the invocation of the
procedure $Full$ on $S_v$. }
\end{center}
\end{figure*}

The Procedure $Full$ accepts as input a star subgraph $S_v$ of a
tree $T$ and transforms it into a full binary tree $\mathcal B_v$
rooted at $v$ that has depth $\lfloor \log (ch(v)+1) \rfloor$, such
that $c_1(v),c_2(v),\ldots,c_{ch(v)}$ are arranged in the resulting
tree by increasing order of level. (See Figure \ref{full} for an
illustration.) Specifically, $c_1(v)$ and $c_2(v)$ become the left
and right children of $v$ in $\mathcal B_v$, respectively. More
generally, for each index $i=1,2,\ldots,\lfloor \frac{ch(v)-1}{2}
\rfloor$, $c_{2i+1}(v)$ and $c_{2i+2}(v)$ become the left and right
children of the vertex $c_i(v)$, respectively.

\begin{figure*}[htp]
\begin{center}
\begin{minipage}{\textwidth} 
\begin{center}
\setlength{\epsfxsize}{6.5in}
\epsfbox{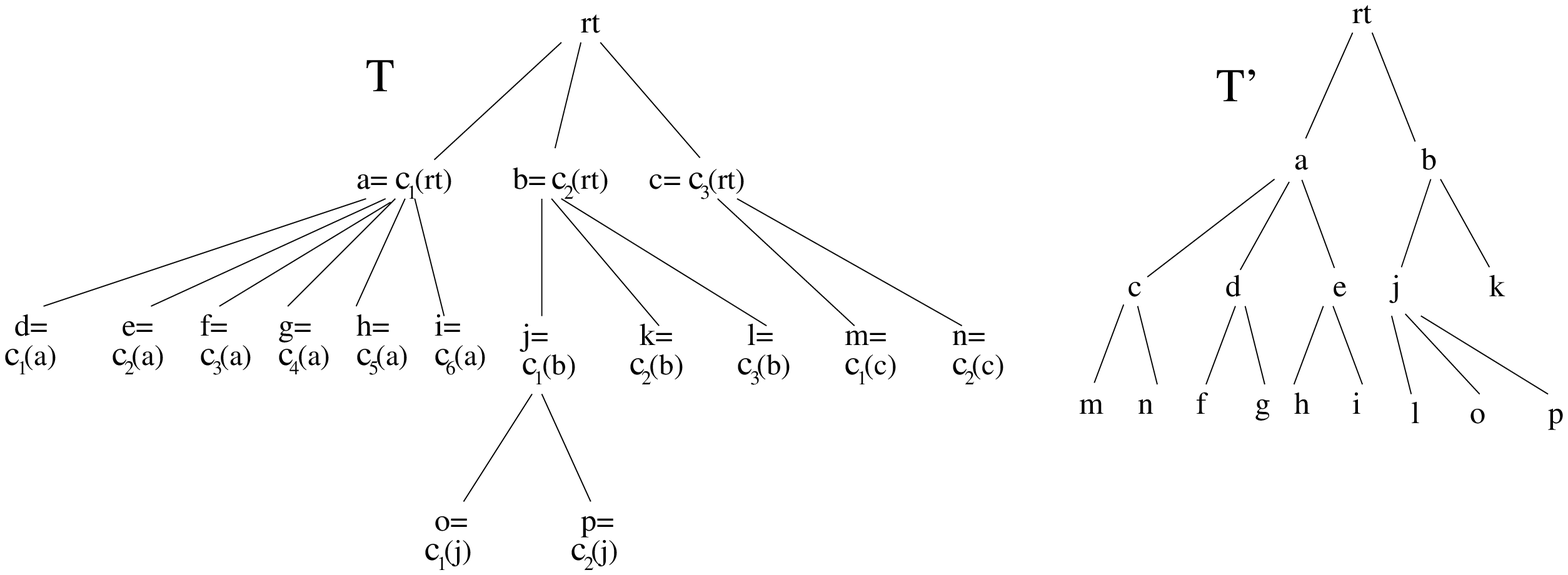}
\end{center}
\end{minipage}
\caption[]{ \label{4Extension} \sf A spanning tree $T$ of
$\{rt,a,b,\ldots,p\}$ rooted at the vertex $rt$ is depicted on the
left.
The 4-ary spanning tree $T'$ of $\{rt,a,b,\ldots,p\}$ rooted at $rt$
is depicted on the right. The tree $T'$ is obtained as a result of
the invocation of the procedure $4Extension$ on $T$. }
\end{center}
\end{figure*}

The Procedure $4Extension$ accepts as input a tree $T$ and transforms
it into a 4-ary tree spanning the original set of vertices.
Basically, the procedure invokes the Procedure $Full$ on every star of
the original tree $T$. More specifically, if the tree $T$ contains
only one node, then the Procedure $4Extenstion$ leaves the tree
intact. Otherwise, it is invoked recursively on each of the subtrees
$T_1,T_2,\ldots,T_{ch(rt)}$ of $T$.
At this point the Procedure
$Full$ is invoked with the parameter $S_{rt}$ and transforms the star
subgraph $S_{rt}$ of the root
 into a full binary tree $\mathcal B_{rt}$ as described above.
 (See Figure \ref{4Extension}
for an illustration.)

It is easy to verify that for any tree $T$, the tree $T'$ obtained
as a result of the invocation of the Procedure $4Extension$ on the
tree $T$ has the same vertex set. Moreover, in the following lemma
we show that no vertex in $T'$ has more than four children.
\begin {lemma} \label{prob1}
$T'$ is a 4-ary tree such that its root $rt$ has at most two
children.
\end {lemma}
\proof The vertices $c_1(rt)$ and $c_2(rt)$ are the only children of
$rt$ in $T'$. For a vertex $v \in V$, $v \ne rt$, let $u$ denote the
parent of $v$ in the original tree $T$. Let $i$ be the index such
that $v = c_i(u)$. If $i \in \{1,2\}$, then $u$ remains the parent
of $v$ in $T'$. Otherwise, $c_{\lfloor (i-1)/2 \rfloor}(u)$ becomes
the parent of $v$ in $T'$. Moreover, $v$ will have at most four
children in $T'$, specifically, $c_{2i+1}(u)$, $c_{2i+2}(u)$,
$c_1(v)$ and $c_2(v)$. If $v$ is a leaf in $T$ then it will have at
most two children in $T'$, $c_{2i+1}(u)$ and $c_{2i+2}(u)$. If $i >
\left\lfloor \frac{ch(u)-1}{2} \right\rfloor$, then $v$ may have
only two children in $T'$, $c_1(v)$ and $c_2(v)$. In this case if
$v$ is a leaf in $T$ then it is a leaf in $T'$ as well.
\QED

{\bf Remark:} Observe that this procedure never increases vertex
degrees, and thus for a binary (respectively, ternary) tree $T$, the
resulting tree $T'$ is binary (resp., ternary) as well.

In the next lemma we show that the height of $T'$ is not much
greater than that of $T$.
\begin {lemma}
$h(T') \leq h(T) + \log \vert T \vert$.
\end {lemma}
\proof
The proof is by induction on $h=h(T)$.
\\\emph{Basse: $h=0$.} The
claim holds vacuously in this case, since  $T' = T$.
\\\emph{Induction Step:} We assume the correctness of the claim
for all trees of depth at most $h-1$, and prove it for trees of
depth $h$. \\Consider a tree $T$ of depth $h$. By the induction
hypothesis, for all indices $i, 1 \leq i \leq ch(rt)$,
\begin {equation} \label {ind}
h(T'_{i}) \leq h(T_{i}) + \log |T_{i}|.
\end {equation}
Note that the root $c_i(rt)$ of $T_i'$ has depth $\lfloor \log (i+1)
\rfloor$ in the tree $T'$. Hence the depth $h(T')$ of $T'$ is given
by
\begin {equation} \label{eq2}
h(T') ~=~ \max \{h(T'_{i}) + \lfloor \log (i+1) \rfloor : i \in
\{1,2,\ldots,ch(rt)\}\}.
\end {equation}
Let $t$ be an index in  $\{1,2,\ldots,ch(rt)\}$ realizing the maximum,
i.e., $h(T') = h(T'_{t}) + \lfloor \log (t+1) \rfloor.$ By equations
(\ref{ind}) and (\ref{eq2}),
$$
h(T') ~\leq~ h(T_{t}) + \log |T_{t}|+ \lfloor \log (t+1) \rfloor.
$$
Note also that $h(T_{t}) \leq h -1$, and $\lfloor \log (t+1) \rfloor
\leq \lfloor \log t \rfloor +1.$ Hence, it holds that $$h(T') ~\le~
h + \log |T_t| + \lfloor \log t \rfloor.$$ Recall that the children
of $rt$ are ordered such that
$$|T_1| ~\ge~ |T_2| ~\ge~ \ldots ~\ge~ |T_{ch(rt)}|.$$
Hence,
$$t \cdot |T_{t}| ~\leq~ \sum_{i=1}^t \vert T_{i} \vert ~\le~ \vert T \vert.$$
Consequently, $\log |T_{t}|+ \lfloor \log t \rfloor \leq \log \vert
T \vert$, and we are done.
\QED

In the following lemma we argue that the weight of the resulting
tree $T'$ is not much greater than the weight of the original tree
$T$.
\begin {lemma} \label{ana}
$\omega(T') \leq 3 \cdot \omega(T)$.
\end {lemma}
\proof
For each vertex $v$ in $T$, its star subgraph $S_{v} = (V_v,E_v)$ is
replaced by a full binary tree $F_v = Full(S_v) = (V_v,E'_v)$. The
weight of $S_{v}$ is given by $$\omega(S_v) = \sum_{i=1}^{ch(v)}
\omega(v,c_i(v))~.$$ Next, we show that the weight of $F_v$ is at
most three times greater than the weight of $S_{v}$. This will imply
the statement of the lemma.

By the triangle inequality, for each edge
 $e=(x,y)$ in $E'_v$, we have $$\omega(e) \le \omega(v,x) + \omega(v,y).$$
Therefore,
$$\omega(F_v) ~=~ \sum_{e \in E'_v}\omega(e) ~\le~ \sum_{e=(x,y) \in
E'_v}\left(\omega(v,x)+\omega(v,y)\right)~.$$ Denote the degree of a
vertex $z$ in $F_v = (V,E'_v)$ by $deg(z,F_v)$. It is easy to verify
by double counting that $$ \sum_{e=(x,y) \in
E'_v}\left(\omega(v,x)+\omega(v,y)\right) ~=~ \sum_{u \in V_v}
deg(u,F_v) \cdot \omega(v,u)~.$$ Since $F_v$ is a binary tree, for
each vertex $z$ in $F_v$, $deg(z,F_v) \leq 3$. Hence,
$$\omega(F_v) ~\le~ 3\cdot \sum_{u \in V_v} \omega(v,u) ~=~ 3\cdot
\sum_{i=1}^{ch(v)} \omega(v,c_i(v)) + 3\cdot \omega(v,v) ~=~ 3\cdot
\omega(S_{v}).$$
\QED

The next corollary summarizes the properties of the Procedure
$4Extension$.
\begin {corollary} \label{main}
For a $\vartheta$-tree $T=(V,E)$, the Procedure $4Extension(T)$
constructs a $4$-ary $\vartheta$-tree $T'=(V,E')$ such that $h(T)
\le h(T') \leq h(T) + \log \vert T \vert$ and $\omega(T') \leq
3\cdot \omega(T)$.
\end {corollary}

 Next, we demonstrate that a 4-ary tree can be transformed to a
binary tree, while increasing the depth and weight only by constant
factors. We did not make any special effort to minimize these
constant factors.

\begin{figure*}[htp]
\begin{center}
\begin{minipage}{\textwidth} 
\begin{center}
\setlength{\epsfxsize}{4.5in}
\epsfbox{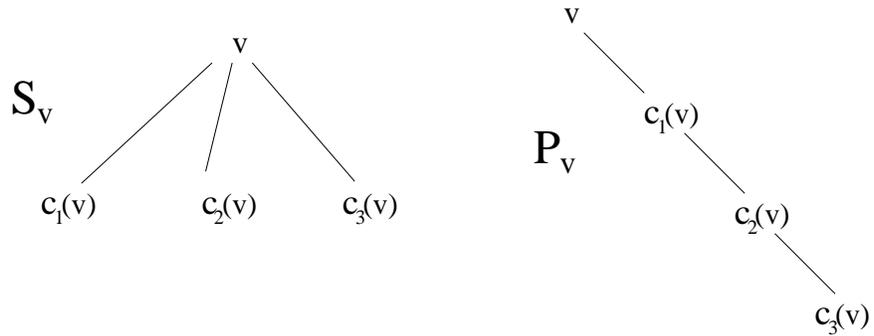}
\end{center}
\end{minipage}
\caption[]{
\label{Path}
\sf
The tree $S_v$ on the left is the star subgraph rooted at $v$ and having the three
vertices $c_1(v),c_2(v),c_3(v)$ as its leaves.
The path $P_v$ on the right is obtained as a result of the invocation
of the procedure $Path$ on $S_v$.
}
\end{center}
\end{figure*}

The Procedure $Path$ accepts as input a star subgraph $S_v$ of a
tree $T$ and transforms it into a path $P_v =
(c_0(v)=v,c_1(v),\ldots,c_{ch(v)}(v))$. (See Figure \ref{Path} for
an illustration.)

\begin{figure*}[htp]
\begin{center}
\begin{minipage}{\textwidth} 
\begin{center}
\setlength{\epsfxsize}{6.5in}
\epsfbox{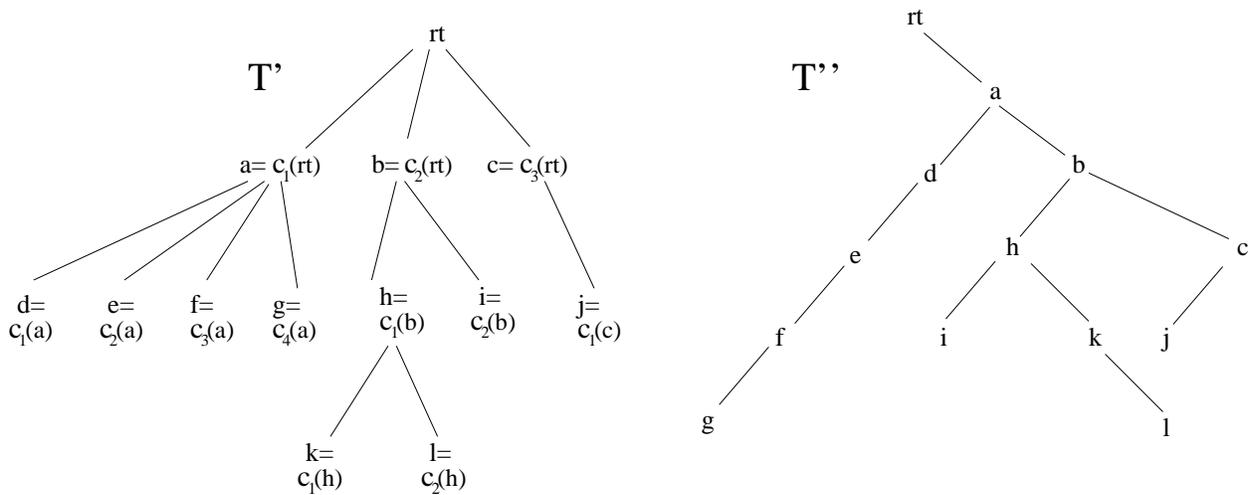}
\end{center}
\end{minipage}
\caption[]{ \label{BinExtension} \sf A 4-ary spanning tree $T'$ of
$\{rt,a,b,\ldots,l\}$ rooted at the vertex $rt$ is depicted on the
left.
The binary spanning tree $T''$ of $\{rt,a,b,\ldots,l\}$ rooted at
$rt$ is depicted on the right. The tree $T''$ is obtained as a
result of the invocation of the procedure $BinExtension$ on $T'$. }
\end{center}
\end{figure*}

The Procedure $BinExtension$ accepts as input a 4-ary tree
$T'=(V,E')$ rooted at a given vertex $rt \in V$ and transforms it
into a binary tree spanning the original set of vertices. If the
tree $T'$ contains only one node, then the Procedure $BinExtenstion$
leaves the tree intact. Otherwise, for every inner vertex $v \in V$,
the procedure replaces the star $S_v = (v,c_1(v),\ldots,c_{ch}(v))$
by the path $P_v = Path(S_v)$. (See Figure \ref{BinExtension} for an
illustration).

In the next lemma we argue that $T'':=BinExtension(T')$ is a binary
tree.
\begin {lemma}
$T''$ is a binary tree such that its root $rt$ has at most one
child.
\end {lemma}
{\bf Remark:} The notation $c_i(v)$ refers to the $i$th child of $v$
in the tree $T'$ that is provided to the Procedure $BinExtension$ as
\emph{input}. \\
\proof
 The vertex $c_1(rt)$ is the only child of $rt$ in $T'$.
 For a vertex $v \in V$, $v \ne rt$, let $u$ denote the parent of
 $v$ in the tree $T'$. Let $i$ be the index such that $v = c_i(u)$.
 If $i=1$ then $u$ is the parent of $v$ in $T''$. Otherwise,
 $c_{i-1}(u)$ is the parent of $v$ in $T''$. Moreover, if $i < ch(u)$
 then $v$ may have two children in $T''$, specifically, $c_{i+1}(u)$
 and $c_1(v)$. If $v$ is a leaf in $T'$ then it will have only one
 child, $c_{i+1}(u)$. If $i = ch(u)$ then $v$ may have at most one
 child in $T''$, specifically, $c_1(v)$. In this case if $v$ is a
 leaf in $T'$ then it is a leaf in $T''$ as well.
\QED

The next two statements imply that both the depth and the weight of
$T'$ and $T''$ are the same, up to a constant factor.
\begin {lemma}
$h(T') \le h(T'') \le 4 \cdot h(T')$.
\end {lemma}
\proof For each vertex $v$ in $T'$, its star subgraph $S_{v}$ is
transformed into a path $Path(S_v)$ of depth $ch(v)$. Since $T'$ is
a 4-ary tree, $ch(v) \le 4$. Thus the ratio between the depth of
$Path(S_v)$ and the depth of $S_v$ is at most $4$. It follows that
for every vertex $v \in V$, the distance between the root and $v$ in
$T''$ is at most four times larger than the distance between them in
$T'$. \QED

\begin {lemma}
$\omega(T'') \le 2 \cdot \omega(T')$.
\end {lemma}
The proof of this lemma is very similar to that of Lemma \ref{ana},
and it is omitted.

We summarize the properties of the reduction from 4-ary to binary
trees in the following corollary.
\begin {corollary} \label{main2}
For a 4-ary $\vartheta$-tree $T'=(V,E')$, the Procedure
$BinExtension(T')$ constructs a binary $\vartheta$-tree $T''$
such that $h(T') \le h(T'') \leq 4 \cdot h(T')$ and
$\omega(T'') \leq 2\cdot \omega(T')$.
\end {corollary}

Corollary \ref{main} and Corollary \ref{main2} imply the following
statement.
\begin {lemma} \label{extension}
For a high $\vartheta$-tree $T=(V,E)$, the invocation
$BinExtension(4Extension(T))$ returns a binary
$\vartheta$-tree $T''$ such that $h(T) \le h(T'') \leq 8 \cdot
h(T)$ and $\omega(T'') \leq 6 \cdot \omega(T)$.
\end {lemma}

We are now ready to prove the desired lower bound for high
$\vartheta$-trees.

\begin {theorem} \label{genasym}
For sufficiently large integers $n$ and $h$, $h \ge \log n$, the
minimum weight $W(n,h)$  of a general $\vartheta$-tree that has
depth at most $h$ is at least $\Omega(k \cdot n)$, for some $k$
satisfying $h = \Omega(k \cdot n^{1/k})$.
\end {theorem}
\proof
Given an arbitrary $\vartheta$-tree $T$ of depth $h(T)$
at least $\log n$, Lemma \ref{extension} implies the existence of a
binary tree $\tilde T$ such that $h(\tilde T) \leq 8 \cdot h(T)$ and
$\omega(\tilde T) \leq 6 \cdot \omega(T)$. By Theorem \ref{asym},
the weight $\omega(\tilde T)$ of $\tilde T$ is at least $\Omega(k
\cdot n)$, for some $k$ satisfying $h(\tilde T) =\Omega(k \cdot
n^{1/k})$. Since $\omega(T) = \Omega(\omega(\tilde T))$ and $h(T) =
\Omega(h(\tilde T))$, it follows that the weight $\omega(T)$ of $T$
is at least $\Omega(k \cdot n)$ as well, and $k$ satisfies the
required relation $h(T) = \Omega(h(\tilde T)) = \Omega(k \cdot
n^{1/k})$.
\QED

Clearly, the lightness of a graph $G$ is at least as large as that
of any BFS tree of $G$, implying the following result.
\begin {corollary}
For  sufficiently large integers $n$ and $h$, $h \ge \log n$, any
spanning subgraph with hop-radius at most $h$ has lightness at least
$\Omega(\Psi)$, for some $\Psi$ satisfying $h = \Omega(\Psi \cdot
n^{1/\Psi})$.
\end {corollary}

\subsection {Lower Bounds for Low Trees} \label{sublowtr}

In this section we devise lower bounds for
the weight of \emph{low} $\vartheta$-trees, that is, trees of depth
$h$ at most logarithmic in $n$.

Our strategy is to show lower bounds for the covering  (Section
\ref{seccov}), and then to translate them into lower bounds for the
weight (Section \ref{secasym}).

\subsubsection {Lower Bounds for Covering} \label{seccov}

 In this
section we prove lower bounds for the covering of trees that have
depth $h \le \frac{1}{5} \cdot \log n$.

The following lemma shows that for any $\vartheta$-tree $T$, the
 covering of $T$ cannot be much smaller than
the maximum degree of a vertex in $T$.
\begin {lemma} \label{degree}
For a tree $T$ and a vertex $v$ in $T$, the covering $\chi(T)$ of
$T$ is at least $(deg(v)-2)/2$.
\end {lemma}
\proof Consider an edge $(v,u)$ in $T$. Since every edge adjacent to
$v$ is either left or right with respect to $v$, by the pigeonhole
principle either at least $\left \lceil \frac{deg(v)}{2} \right
\rceil$ edges are left with respect to $v$ or at least $\left \lceil
\frac{deg(v)}{2} \right \rceil$ edges are right with respect to $v$.
Suppose without loss of generality that at least $\left \lceil
\frac{deg(v)}{2} \right \rceil$ edges are right with respect to $v$,
and denote by $\mathcal R$ the set of edges which are adjacent to
$v$ and right with respect to $v$. Note that all these edges except
possibly the edge $(v,v+1)$ cover the vertex $(v+1)$, and thus the
covering of $(v+1)$ is at least $$|\mathcal R| -1 ~\ge~ \left \lceil
\frac{deg(v)}{2} \right \rceil -1 \ge  \frac{deg(v) -2}{2}.$$
\QED

Though the ultimate goal of this section is to establish a lower
bound for the range $h \le \frac{1}{5} \cdot \log n$, our argument
provides a relationship between covering and depth $h$ in a wider
range $h \le \ln n$.
The next lemma takes care of the simple case of $h$ close to $\ln
n$. The far more complex range of small values of $h$ is analyzed in
Lemma \ref{prop}.

\begin {lemma} \label{ext}
For a tree  that has depth $h$ and covering $\chi$, such that
$\frac{\ln n}{2} < h \le {\ln n}$, $$\chi ~>~ \frac{1}{e^2} \cdot h
\cdot n^{1/h} -h.$$
\end {lemma}
\proof
We claim that for each $\frac{\ln n}{2} < h \le
{\ln n}$, the right hand-side is of a negative value. To see this,
note that in the range $[1,\ln n]$, the function $g(h) =
\frac{1}{e^2} \cdot h \cdot n^{1/h} -h$ is monotone decreasing with
$h$. Thus, $g(h)$ is smaller than $g(\frac{\ln n}{2}) = 0$ in the
range $(\frac{\ln n}{2},\ln n]$, as required.
\QED

\begin {lemma} \label{prop}
For a tree that has depth $h$ and covering $\chi$, such that $h \le
\frac{\ln n}{2}$,
$$\chi
> \frac{1}{10} \cdot h \cdot n^{1/h} -h.$$
\end {lemma}
\proof
\\The proof is by induction on $h$, for all values of $n \geq e^{2h}$.
\\\emph{Base:} The statement
holds vacuously for $h=0$. For
$h=1$, the degree of the root is
$n-1$. Hence Lemma \ref{degree} implies that the covering is at least
$\frac{n-3}{2} \ge {\frac{1}{10}\cdot n}-1$, for any $n
\ge e^{2}$.
\\\emph{Induction Step:} We assume the claim for all trees of depth at
most $h$, and prove it for trees of depth $h+1$. Let $T$ be a tree
on $n$ vertices that has depth $h+1$ and covering $\chi$, such that
$h+1 \le \frac{\ln n}2$. If $deg(rt) \ge \sqrt n$, then Lemma
\ref{degree} implies that the covering is at least $\frac{\sqrt n
-2}{2}$. It is easy to verify that for $h \ge 1$ and $n \ge
e^{2(h+1)}$, $$\frac{\sqrt n -2}{2} \ge \frac{1}{10} \cdot (h+1)
\cdot n^{1/(h+1)} -(h+1).$$ Henceforth, we assume that $deg(rt) <
\sqrt n$.

For a child $u$ of $rt$, denote by $T_u$ the subtree of $T$ rooted
at $u$. Such a subtree $T_u$ will be called \emph{light} if $|T_u| <
e^h$. Denote the set of all light subtrees of $T$ by $\mathcal{L}$,
and define $T'$ as the tree obtained from $T$ by omitting all light
subtrees from it, namely, ${\displaystyle T' = T ~\setminus
~\bigcup_{T_u \in \mathcal{L}} T_u}$. Observe that the covering
$\chi'$ of $T'$ is at most $\chi$. Next, we obtain a lower bound on
$\chi'$.

Observe that $$\left|\bigcup_{T_u \in \mathcal{L}} T_u\right| ~=~
\sum_{T_u \in \mathcal{L}} |T_u| ~<~ deg(rt) \cdot e^h ~<~ \sqrt n
\cdot \frac{\sqrt n}{e} ~=~ \frac{n}{e},$$ implying that
\begin{equation} \label{ntag}
n' ~=~ |T'| ~=~ |T| - \left|\bigcup_{T_u \in \mathcal{L}} T_u\right| ~>~ n-n/e.
\end{equation}
Denote the subtrees of $T'$ by
$T_1,\ldots,T_k$, where $k \leq deg(rt)$,
 and define for each $1 \le i \le k$, $n_i = |T_i|$, $\chi_i = \chi(T_i)$,
 and $h_i = depth(T_i)$.
 Fix an index $i$, $i \in \{1,\ldots,k\}$.
 Observe that $n_i \ge e^h$, or equivalently, $h \le \ln n_i$.
 Since $h_i \le h$,
we have that $h_i \le \ln n_i$. By the induction hypothesis and Lemma \ref{ext},
$$\chi_i ~>~ \frac{1}{10} \cdot
h_i \cdot n_i^{1/h_i} - h_i.$$ As the function $f(x) = \frac{1}{10}
\cdot x\cdot n_i^{1/x} -x$ is monotone decreasing in the range
$[1,\ln n_i]$, and since $h_i \le h \le \ln n_i$, it follows that
$$\chi_i ~>~ \frac{1}{10} \cdot
h_i \cdot n_i^{1/h_i} - h_i ~\ge~ \frac{1}{10} \cdot h \cdot n_i^{1/h}
-h.$$ Therefore, for each index $i$, $1 \le i \le k$,
\begin {equation} \label{fi}
 n_i ~<~ \left(\frac{\chi_i+h}{\frac{1}{10} \cdot h}\right)^h.
\end {equation}
For each $i$, $1 \le i \le k$, choose some vertex $v^*_i$ in $T_i$
with maximum covering, i.e., $\chi_{T_i}(v^*_i)=\chi_i$. We assume
without loss of generality that $$v^*_{1}~<~v^*_{2}~<~ \ldots ~<~
v^*_{k}.$$ (Note that this order is not necessarily the order of the
children $u_1,u_2,\ldots,u_{ch(rt)}$ of the root of $T$. In other
words, it may be the case that $v^*_1 < v^*_2$ but $u_1 > u_2$.)
Let $p$ be the index for which
$v^*_{p} < rt$ and $v^*_{p+1}
> rt$.

\begin {claim}
$\chi' \ge \max\{\chi_{1},\chi_{2}+1,\ldots,\chi_{p}+(p-1)\}$.
\end {claim}
\proof
Consider the path in $T'$ between $rt$ and $v^*_{i}$, for each $1
\le i \le p-1$. For each index $j$, $i+1 \le j \le p$, the vertex
$v^*_j$ must be covered by at least one edge in that path. It
follows that $\chi_{T'}(v^*_{j}) \geq \chi_{j} + j-1$, for each $1
\leq j \leq p$, as required.
\QED

A symmetric argument yields the following inequality.
\begin {claim} \label{cl2}
$\chi' \ge \max\{\chi_{k}, \chi_{k-1}+1, \ldots,
\chi_{p+1}+(k-p-1)\}$.
\end {claim}

Suppose without loss of generality that ${\displaystyle \sum_{i=1}^p
n_i \ge \sum_{i=p+1}^k n_i}$. (The argument is symmetric if this is
not the case.) Since ${\displaystyle n'= |T'|= \sum_{i=1}^k n_i,}$
it follows that ${\displaystyle \frac{n'}{2} ~\le~ \sum_{i=1}^p
n_i.}$ By (\ref{fi}), $ \sum_{i=1}^p n_i < \sum_{i=1}^p
\left(\frac{\chi_i + h}{\frac{1}{10}\cdot h}\right)^h.$ Notice that
for each $1 \le i \le p$, $\chi_i +i-1 ~\le~ \chi'.$ Consequently,
\begin{eqnarray*}
\sum_{i=1}^p \left(\frac{\chi_i + h}{\frac{1}{10}\cdot h}\right)^h
&\le& \sum_{i=1}^p \left(\frac{\chi' + h -(i-1)}{\frac{1}{10}\cdot
h}\right)^h \cr\cr \\ &=& 10^h \cdot \left(\frac{1}{h}\right)^h
\cdot \sum_{i=1}^p \left(\chi' + h -(i-1)\right)^h.
\end{eqnarray*}

Since $\sum_{i=1}^p \left(\chi' + h -(i-1)\right)^h <
\frac{\left(\chi'+h+1\right)^{h+1}}{h+1},$ we conclude that
\begin {equation} \label{rr}
\begin{array}{ll}
 {\displaystyle \frac{n'}{2} ~<~ {10}^h
\cdot \left(\frac{1}{h}\right)^h \cdot
\frac{\left(\chi'+h+1\right)^{h+1}}{h+1}} \cr\cr {\displaystyle
~~~~=~ \frac{1}{10} \cdot \left(\frac{h+1}{h}\right)^h \cdot
\left(\frac{\chi'+h+1}{\frac{1}{10}\cdot (h+1)}\right)^{h+1} ~\le~
\frac{e}{10} \cdot \left(\frac{\chi'+h+1}{\frac{1}{10}\cdot
(h+1)}\right)^{h+1}}.
\end{array}
\end {equation}

By (\ref{ntag}), $n-n/e < n'$, and thus (\ref{rr}) implies that
$$\frac{n\cdot (1-1/e)}{2} ~<~ \frac{e}{10} \cdot
\left(\frac{\chi'+h+1}{\frac{1}{10}\cdot (h+1)}\right)^{h+1},$$ and
consequently, $n  <
\left(\frac{\chi'+h+1}{\frac{1}{10}\cdot(h+1)}\right)^{h+1}.$ It
follows that $\chi \ge \chi' > \frac{1}{10} \cdot (h+1) \cdot
n^{\frac{1}{h+1}} -(h+1),$ as required.
\QED

The next corollary follows easily from the above analysis.
\begin {corollary} \label{work}
For a tree that has depth $h$ and covering $\chi$, such that $h \le
\frac{1}{5} \cdot \log n$, it holds that $\chi > \frac{1}{20} \cdot
h \cdot n^{1/h}.$
\end {corollary}
\proof
It is easy to verify that for any $h \le \frac{1}{5} \cdot \log n$,
it holds that $h < \frac{1}{2} \cdot \left(\frac{1}{10} \cdot h
\cdot n^{1/h}\right).$ Hence, the statement follows from Lemma
\ref{prop}.
\QED

\subsubsection {Lower Bounds for Weight} \label{secasym}

In this section we employ the lower bound for the covering
(Corollary \ref{work}) to show lower bounds for the weight of
$\vartheta$-trees of depth $h < \log n$. For the range $h \le
\frac{1}{10} \cdot \log n$, we employ a technique due to Agarwal et
al. \cite{AWY05} to translate the lower bound on the covering
established in the previous section into the desired lower bound for
high trees. (Our lower bound on the covering is, however,
significantly stronger than that of \cite{AWY05}.) Somewhat
surprisingly, our lower bound for the complementary range
$\frac{1}{10} \cdot \log n < h < \log n$ relies on our lower bound
for the range $h \ge \log n$ (Theorem \ref{genasym}).

The following claim establishes a relation between the weight of a tree, and the sum of
coverings of its vertices.
\begin {lemma} \label{covwt}
For any $\vartheta$-tree $T$ of depth $h$, it holds that
$$\sum_{e \in E(T)} \omega(e) ~=~ \sum_{v \in V(T)} \chi(v) + n-1.$$
\end {lemma}
\proof
For an edge $e \in E(T)$, let $\varphi(e)$ denote the number of
vertices covered by $e$. Clearly, $\omega(e) = \varphi(e) + 1$. It
is easy to verify by double counting that
$$\sum_{v \in V(T)} \chi(v) ~=~ \sum_{e \in E(T)} \varphi(e).$$
Consequently,
$$\sum_{e \in E(T)} \omega(e) =  \sum_{e \in E(T)}
(\varphi(e)+1) = \sum_{v \in V(T)} \chi(v) + n-1.$$
\QED

 Denote the minimum covering of a $\vartheta$-tree of depth \emph{at most} $h$ by $\tilde
\chi(n,h)$. Since the sequence $\left(\chi(n,h)\right)_{h=1}^{n-1}$
is monotone non-increasing (by Lemma \ref{moncov}), it holds that
\begin{equation} \label{usemon} \tilde \chi(n,h) = \chi(n,h).
\end {equation}
 The proof of the following lemma is
closely related to the proof of Lemma 2.1 from \cite{AWY05}, and is
provided for completeness.
\begin {lemma}
For a sufficiently large integer $n$, and a positive integer $h$, $h
\le \frac{1}{10} \cdot \log n$, it holds that
$$W(n,h) = \Omega(n \cdot h \cdot n^{1/h}).$$
\end {lemma}
\proof
Let $T$ be a $\vartheta$-tree of minimum weight $W(n,h)$, that has
depth at most $h$, $h \le \frac{1}{10} \cdot \log n$, and covering
$\chi$. Denote the root vertex of $T$ by $rt$. Let $g(n,h) =
\frac{1}{20} \cdot h \cdot n^{1/h}$. By Corollary \ref{work} and
(\ref{usemon}),
\begin {equation} \label{rnh}
\chi > \tilde \chi(n,h) \ge g(n,h).
\end {equation}
A vertex $v$ is called \emph{heavy} if $\chi(v) \ge \frac{1}{6}
\cdot g(n,h)$, and it is called \emph{light} otherwise. Let
$\mathcal H$ be the set of heavy vertices in $T$. Next, we prove
that $|\mathcal H| \ge n/2$. Since by Lemma \ref{covwt},
$$\sum_{e \in E(T)} \omega(e) ~\ge~ \sum_{v \in V(T)} \chi(v) ~\ge~ |\mathcal H| \cdot \frac{1}{6} \cdot g(n,h),$$
this would complete the proof of the lemma.

Suppose to the contrary that $|\mathcal H| < n/2$. We need the
following definition. A set $B = \{i,i+1,\ldots,j\}$ of consecutive
vertices is said to be a \emph{block} if $B \subseteq \mathcal H$.
If $(i-1) \ne B$ and $(j+1) \ne B$ then $B$ is called a
\emph{maximal block}. Decompose $\mathcal H$ into a set of
(disjoint) maximal blocks $\mathcal B = \{B_1,B_2,\ldots,B_k\}$.
By contracting the
induced subgraph of each $B_i$ into a single node $w_i$, for $1 \le
i \le k$, we obtain a new graph $G'=(V',E')$. Observe that $G'$ is
not necessarily a tree and may contain multiple edges connecting the same pair of
vertices. For each pair of
vertices $u$ and $v$ in $V'$, omit all edges except for the lightest
one that connect $u$ and $v$ in $T'$. If $rt \in w_i$, for some
index $i \in [1,k]$, then designate $w_i$ as the new root vertex
$rt$. Let $T' = (V',E'')$ be the BFS tree of $G'$
rooted at $rt$. Clearly, $n'=|V'| \ge |V
\setminus \mathcal H| > n/2$, and $h(T') \le h(T) =h$. Let
$v_1,v_2,\ldots,v_{n'}$ be the vertices of $T'$ in an increasing
order. Transform $T'$ into a spanning tree $\tilde T$ of
$\vartheta_{n'}$ by mapping each $v_i$ to the point $i$ on the $x$-axis, for
$i=1,2,\ldots,n'$.

\begin {claim} \label{covttag}
$\chi(\tilde T) \le \frac{1}{3} \cdot g(n,h) + 1$.
\end {claim}
\proof
First, observe that all light vertices in $T$ remain light in
$\tilde T$. For a contracted heavy vertex $w_i$, we argue that
$\chi(w_i) \le \chi(w_i^-) + \chi(w_i^+) + 1$, where $w_i^-$
(respectively, $w_i^+$) is the vertex to the immediate left (resp.,
right) of $w_i$. To see this, note that for each edge $e \ne
(w_i^+,w_i^-)$ that covers $w_i$, $e$ covers either $w_i^-$ or
$w_i^+$.
Since both vertices $w_i^+$ and $w_i^-$ are light, it follows that
 $\chi(w_i) \le 2\cdot
\left(\frac{1}{6} \cdot g(n,h)\right) +1$, and we are done.
\QED

Observe that for $n \ge 4$, we have that $\frac{1}{10} \cdot \log n
\le \frac{1}{5} \cdot \log (n/2).$ Since $h \le \frac{1}{10} \cdot
\log n$, and $n'
> n/2$, it follows that $h \le \frac{1}{5} \cdot \log n'$.
Hence by Corollary \ref{work} and (\ref{rnh}),
\begin{eqnarray*} \chi(\tilde T) &\ge& \tilde \chi(n',h) ~\ge~ g(n',h) ~=~ \frac{1}{20} \cdot h \cdot
{n'}^{\frac 1 h} \\ &>& \frac{1}{20} \cdot h \cdot {(n/2)}^{\frac 1
h} ~=~ \left(\frac{1}{2}\right)^{\frac 1 h} \cdot g(n,h).\end
{eqnarray*} However, for sufficiently large $n$,
$$\left(\frac{1}{2}\right)^{\frac 1 h} \cdot g(n,h) > \frac{1}{3}
\cdot g(n,h) + 1,$$ contradicting Claim \ref{covttag}.
\QED

\begin {lemma} \label{borderl}
For a sufficiently large integer $n$, and a positive integer
$h$, $\frac{1}{10} \cdot \log n \le h < \log n$, it holds that
$W(n,h) = \Omega(n \cdot h \cdot n^{1/h}).$
\end {lemma}
\proof
By Theorem \ref{genasym}, the minimum weight $W(n,\log n)$ of a
$\vartheta_n$-tree that has depth $h$, $h \le \log n$,
is at least $\Omega(k \cdot n)$, for some $k$ satisfying $\log n =
\Omega(k \cdot n^{1/k})$. It follows that $W(n,\log n) = \Omega(\log
n \cdot n)$.
\\ Since by Lemma
\ref{monwt}, the sequence $\left(W(n,h)\right)|_{h=1}^{n-1}$ is monotone non increasing,
it holds that for $h \in [\frac{1}{10} \cdot \log n
,\log n]$,
$$W(n,h) = \Omega(\log n \cdot n) = \Omega(n \cdot h \cdot
n^{1/h}).$$
\QED

Lemmas \ref{covwt} and \ref{borderl} imply the following
lower bound
for $h < \log n$.
\begin {corollary} \label{lowasym}
For a sufficiently large integer $n$, and a positive integer
$h$, $h < \log n$, it holds that $W(n,h) = \Omega(n \cdot h \cdot n^{1/h}).$
\end {corollary}

Clearly, the lightness of a graph $G$ is at least as large as that
of any BFS tree of $G$, implying the following result.
\begin {theorem}
For a sufficiently large integer $n$ and a positive integer $h$, $h <
\log n$, any spanning subgraph with hop-radius at most $h$ has
lightness at least $\Psi = \Omega(h \cdot n^{1/h})$.
\end {theorem}

\section{Euclidean Spanners} \label{sec:euc}

The following theorem is a direct corollary of Theorem \ref{genasym}
and Corollary \ref{lowasym}. It implies that no construction that
provides Euclidean spanner with hop-diameter $O(\log n)$ and
lightness $o(\log n)$, or vice versa, is possible. This settles the
open problem of \cite{ADMSS95,AWY05}.

\begin {theorem}
\label{settle} For a sufficiently large integer $n$, any spanning
subgraph of $\vartheta$ with hop-diameter at most $O(\log n)$ has
lightness at least $\Omega(\log n)$, and vice versa.
\end {theorem}

{
\proof
First, we show that any $\vartheta$-tree that has depth
at most $O(\log n)$ has lightness at least $\Omega(\log n)$. If the
depth $h$ is at least $\log n$, Theorem \ref{genasym} implies that
the lightness is at least $\Omega(k)$, for some $k$ satisfying $h =
\Omega(k \cdot n^{1/k})$. Observe that for $h = O(\log n)$, any $k$
that satisfies $h = \Omega(k \cdot n^{1/k})$ is at least
$\Omega(\log n)$, as required. By the monotonicity (Lemma
\ref{monwt}), the lower bound of $\Omega(\log n)$ applies for all
smaller values of $h$.

Consider a spanning subgraph $G$ of $\vartheta$ that has hop-diameter
$\Lambda = O(\log n)$. Consider a BFS tree $T$ rooted at some vertex
$rt$.
Obviously, $h(T,rt) \le \Lambda = O(\log n)$, and thus
$\omega(T) = \Omega(\log n) \cdot \omega(MST(\vartheta)) = \Omega(n
\log n)$. Since $\omega(G) \ge \omega(T)$, it follows that the
lightness of $G$ is $\Omega(\log n)$ as well.

Next, we argue that any $\vartheta$-tree that has lightness at most
$O(\log n)$ has depth at least $\Omega(\log n)$. Indeed, by
Corollary \ref{lowasym}, any $\vartheta$-tree of depth $h = o(\log
n)$ has lightness at least $\Omega(h \cdot n^{1/h}) = \omega(\log
n)$. Moreover, if a spanning subgraph $G$ of $\vartheta$ has
lightness $\Psi(G) = O(\log n)$, then its BFS tree $T$ satisfies
$\Psi(T) = O(\log n)$ as well. Therefore, $h(T,rt) = \Omega(\log
n)$, and thus $\Lambda(G) \ge h(T,rt) = \Omega(\log n)$.
\QED
}

\section {Upper Bounds}
\label{sec:ub}

  In this section we devise an upper bound for
 the tradeoff between
 various parameters of
binary
 LLTs. This upper bound is tight up to
 constant factors in the entire range of parameters.

Consider a general $n$-point metric space $M$. Let $T^*$ be an MST
for $M$, and $D$ be an in-order traversal of $T^*$, starting at an
arbitrary vertex $v$. For every vertex $x$, remove from $D$ all
occurrences of $x$ except for the first one. It is well-known (\cite
{CLRS90}, ch. 36) that this way we obtain a Hamiltonian path  $L =
L(T)$ of $M$ of total weight
$$\omega(L) = \sum_{e \in L} \omega(e)
\le 2 \cdot \omega(MST(M)).$$
%
Let $(v_1,v_2,\ldots,v_n) = L$ be the order in which the points of
$M$ appear in $L$. Consider an edge $e'=(v_i,v_j)$ connecting two
arbitrary points in $M$, and an edge $e=(v_q,v_{q+1}) \in E(L)$, $q
\in [n-1].$ The edge $e'$ is said to \emph{cover $e$ with respect to
$L$} if $i \le q < q+1 \le j$. When $L$ is clear from the context,
we write that $e'$ \emph{covers} $e$.

\begin{figure*}[htp]
\begin{center}
\begin{minipage}{\textwidth} 
\begin{center}
\setlength{\epsfxsize}{4.5in} \epsfbox{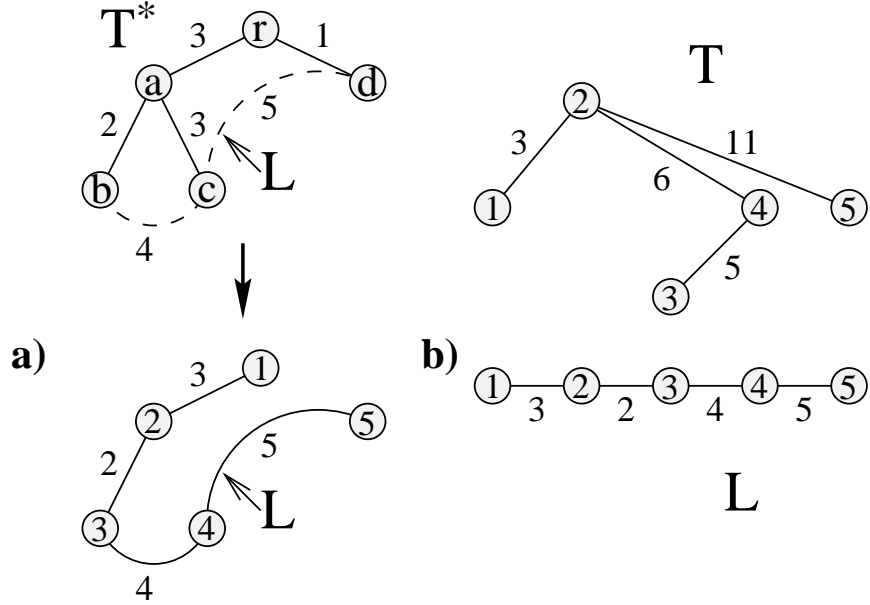}
\end{center}
\end{minipage}
\caption[]{ \label{fig:th} \sf a) The construction of the
Hamiltonian path $L$. b) The load of $T$ on the
 edge $(3,4)$
is 3. }
\end{center}
\end{figure*}

For a spanning tree $T$ of $M$, the number of edges $e' \in E(T)$
that cover an edge $e$ of $E(L)$ is called the \emph{load of $e$ by
$T$} and it is denoted $\xi(e) = \xi_{T}(e)$. The \emph{load of the
tree $T$}, $\xi(T)$, is the maximum load of an edge $e \in E(L)$ by
$T$, i.e.,
$$\xi(T) = \max \{\xi_T(e) ~\vert~ e \in E(L)
\}.$$
Observe that
$$\omega(T) ~\le~ \sum_{e \in L(T)} \xi_T(e) \cdot
\omega(e) ~\le~ \xi(T) \cdot w(L),$$
 and so,
\begin{equation}
\label{xipsi} \xi(T) \ge \frac{\omega(T)}{\omega(L)} \ge \frac{1}{2}
\cdot \Psi(T).
\end {equation}
(See Figure \ref{fig:th} for an illustration.)

In the sequel we provide an upper bound for the load $\xi(T)$ of a
tree $T$, which yields the same upper bound for the lightness
$\Psi(T)$ of $T$, up to a factor of 2. Since the weight of a tree
$T$ does not affect its load, we may assume without loss of
generality that $v_i$ is located in the point $i$ on the $x$-axis.
\begin {lemma} \label{vartogen}
Suppose there exists a $\vartheta$-tree $T$ with load $\xi(T)$ and
depth $h$. Then there exists a spanning tree $T'$ for $M$ with
the same load (with respect to $L(T)$) and depth.
\end {lemma}
{
\proof
The edge set $E'$ of the tree $T'$ is defined by $E' = \{(v_i,v_j)
\vert (i,j) \in T\}$. It is easy to verify that its depth and load
are equal to those of $T$.
\QED
}

Consequently, the problem of providing upper bounds for  general
metric spaces reduces to the problem of providing upper bounds for
$\vartheta$.
%
\subsection {Upper Bounds for High Trees} \label{subhigh}
In this section we devise a construction of $\vartheta_n$-trees with
depth $h \ge \log n$.
%
This construction exhibits tight up to constant factors tradeoff
between load and depth, when the tree depth is at least logarithmic
in the number of vertices. In addition, the constructed trees are
\emph{binary}, and so their maximum degree is \emph{optimal}.

We start with defining a certain composition of binary trees. Let
$n'$ and $n''$ be two positive integers, $n = n'+n''$. Let
$\vartheta'$, $\vartheta''$, and $\vartheta$ be the $n'$-, $n''$-
and $n$-point metric spaces $\vartheta_{n'}$, $\vartheta_{n''}$, and
$\vartheta_{n}$, respectively. Also, let
$\{v'_1,v'_2,\ldots,v'_{n'}\}$, $\{v''_1,v''_2,\ldots,v''_{n''}\}$,
and $\{\tilde v_1,\tilde v_2,\ldots,\tilde v_n\}$ denote the set of
points of $\vartheta'$, $\vartheta''$, and $\vartheta$,
respectively. Consider spanning trees $T'$ and $T''$ for
$\vartheta'$ and $\vartheta''$, respectively. Let $v'=v'_i$ be a
vertex of $T'$. Consider a tree $\tilde T$ that spans the vertex set
$\{\tilde v_1,\tilde v_2, \ldots,\tilde v_{n'+n''}\}$ formed out of
the trees $T'$ and $T''$ in the following way. The root $rt''$ of
$T''$ is added as a right child of $v'$ in $\tilde T$. The vertices
$v'_1,v'_2,\ldots,v'_i=v'$ of $T'$ retain their indices, and are
translated into vertices $\tilde v_1,\tilde v_2,\ldots,\tilde v_i$
in $\tilde T$. The vertices $v''_1,v''_2,\ldots,v''_{n''}$ of $T''$
get the index $i$ of $v'_i$, added to their indices, and are
translated into vertices $\tilde v_{i+1},\tilde v_{i+2},\ldots,
\tilde v_{i+n''}$ in $\tilde T$. Finally, the vertices
$v'_{i+1},v'_{i+2},\ldots,v'_{n'}$ of $T'$ get the number $n''$ of
vertices of $T''$ added to their indices, and are translated into
vertices $\tilde v_{i+1+n''}, \tilde v_{i+2+n''},\ldots, \tilde
v_{n'+n''}$. We say that the tree $\tilde T$ is composed \emph{by
adding $T''$ as a right subtree to the vertex $v$ in $T'$}. Adding a
left subtree $T''$ to a vertex $v$ in $T''$ is defined analogously.

Consider a family of binary $\vartheta$-trees $T(\xi,h)$ with $n =
N(\xi,h)$ vertices, load $\xi$ and depth $h$, $h \ge \xi-1$, $\xi
\ge 1$. These trees are all rooted at the point $1$. For $\xi=1$,
and $h=0$, the tree $T(1,0)$ is a singleton vertex, and so
$N(1,0)=1$. For convenience, we define the load of $T(1,0)$ to be 1.
For $\xi=1$ and $h \ge 1$, the tree $T(1,h)$ is the path $P_{h+1} =
(v_1,v_2,\ldots,v_{h+1})$. The depth of $T(1,h)$ is equal to $h$,
and its load is 1. Hence $N(1,h) = h+1$.

\begin{figure*}[htp]
\begin{center}
\begin{minipage}{\textwidth} 
\begin{center}
\setlength{\epsfxsize}{4.5in} \epsfbox{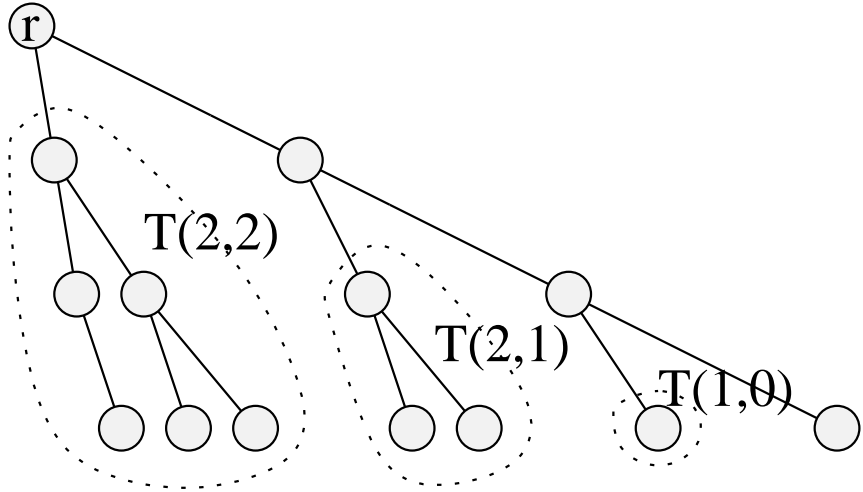}
\end{center}
\end{minipage}
\caption[]{ \label{fig:T} \sf The tree $T(3,3)$. }
\end{center}
\end{figure*}

For $\xi \ge 2$ and $h \ge \xi-1$, the tree $T(\xi,h)$ is
constructed as follows. Let $T'=P_{h+1}$ be the path
$(v_1,v_2,\ldots,v_{h+1})$, and for each index $i$, $i \in [h]$,
define for technical convenience $v'_i = v_i$. For each index $i$,
$i \in [h]$, let $T''_i$ be the tree $T(\xi''_i,h''_i)$, with
$\xi''_i = \min \{\xi-1,h-i+1\}$, $h''_i=h-i$. Observe that for each
$i \in [h]$, $h''_i \ge \xi''_i-1$, and thus the tree $T''_i$ is
well-defined. For every $i \in [h]$, we add the tree $T''_i$ as a
right subtree of $v'_i$ in $T'$.
(See Figure \ref{fig:T} for an illustration.)

\begin {lemma}
\label{lm:depth}
The depth of the resulting tree  $T(\xi,h)$ with respect to the
vertex $v'_1$ is $h$, and its load is $\xi$.
\end {lemma}
{
\proof
The proof is by induction on $\xi$, for all values of $h \ge \xi-1$.
\\\emph{Base $(\xi=1)$:} The tree $T(1,h)$ has load 1 and depth $h$, as required.
\\\emph{Induction Step:} We assume the correctness of the statement for all smaller values
of $\xi$, and prove it for $\xi$. First, the vertex $v'_{h+1}$ is
connected to the root $v'_1$ via the path $P_{h+1}$, and thus the
depth $h(T(\xi,h))$ of $T(\xi,h)$ is at least the length of this
path, that is, $h$. Consider a vertex $u \in V(T''_i)$, for some
$i\in [h]$. The path $P_u$ connecting $v'_1$ and $u$ in $T(\xi,h)$
starts with a subpath $(v'_1, v'_2, \ldots,  v'_i, rt(T''_i))$, and
continues with the unique path $P''_u$ connecting between the root
of $T''_i$ and the vertex $u$ in the tree $T''_i$. The length of the
latter path is no greater than the depth of $T''_i$, which, by the
induction hypothesis, is equal to $h-i$. Hence $|P_u| = i + |P''_u|
\le h$. Hence $h(T(\xi,h)) = h$.

To analyze the load $\xi(T(\xi,h))$ of the tree $T(\xi,h)$, consider
some edge $e=(v_q,v_{q+1})$ of the path $P_{n}$, where $n$ is the
number of vertices in $T(\xi,h)$. The path $T'$ contributes at most
one unit to the load of $e$. In addition, there exists at most one
index $i$, $i \in [h]$, such that the tree $T''_i$ covers the edge
$e$. By induction, the load $\xi(T''_i)$ is equal to $\min
\{\xi-1,h-i+1\}$. Consequently, the load of $e$ is
\begin{eqnarray*} \xi_{T(\xi,h)}(e) &\le& \max \{\min\{
\xi-1,h-i+1\}+1 ~\vert~ i \in [h]\} \\ &=& \min\{ \xi-1,h\}+1 ~=~
\xi.
\end {eqnarray*}
\QED
}

Finally, we analyze the number of vertices $N(\xi,h)$ in the tree
$T(\xi,h)$. By construction,
$$N(\xi,h) = h+1+\sum_{i=1}^h N(\min\{\xi-1,h-i+1\},h-i).$$

\begin {lemma} \label{choosing}
For $h \ge \xi-1$, $N(\xi,h) \ge {h \choose \xi}$.
\end {lemma}
\proof
The proof is by induction on $\xi$. \\\emph{Base:} For $\xi=1$,
$N(1,h) = h+1 \ge {h \choose 1}$, as required. \\\emph{Step:} For
$\xi \ge 2$, and $i \le h- \xi+1$, $\min\{\xi-1,h-i+1\} = \xi-1$.
Hence
\begin{eqnarray*}
N(\xi,h) &\ge& h+1 + \sum_{i=1}^h N(\min\{\xi-1,h-i+1\},h-i)
\\ &\ge& \sum_{i=1}^{h-\xi+1} N(\min\{\xi-1,h-i+1\},h-i)
  ~=~ \sum_{i=1}^{h-\xi+1} N(\xi-1,h-i).
\end {eqnarray*}

Observe that for each index $i$, $i \in [h-\xi+1]$, $h-i  \ge \xi
-1$. Hence, by the induction hypothesis and Fact \ref{fact23}, the
latter sum is at least
$$\sum_{i=1}^{h-\xi+1} {h-i \choose \xi -1} ~=~ {h \choose
\xi}.$$
\QED

The following theorem summarizes the properties of the trees
$T(\xi,h)$.
\begin {theorem} \label{htwolog}
For a sufficiently large $n$, and $h$, $h \ge  2 \log n $, there
exists a binary $\vartheta_n$-tree that has depth at most $h$ and
load at most $\xi$,
where $\xi$ satisfies ($h = O(n^{1/\xi} \cdot \xi)$ and $\xi =
O(\log n)$).
\end {theorem}
{
\proof
First, note that for $h$ in this range,  ${h \choose \log n} > n$.
By Lemma \ref{choosing},
 for any pair of positive integers $h$ and $\xi$ such that $\xi-1 \le h$,
it holds that $N(\xi,h) \ge {h\choose \xi}$. Further, observe that
$N(\xi,h)$ is monotone increasing with both $\xi$ and $h$, in the
entire range $0 \le \xi-1 \le h$. It follows that for $h \ge 2 \log
n$, there exists a positive integer $\xi$, $\xi \le h$, such that
$N(\xi,h-1) \le n < N(\xi,h)$.

Consider the binary tree obtained from the tree $T(\xi,h)$ by
removing $N(\xi,h)-n$ leaves from it, one after another. Clearly,
the resulting tree has depth at most $h$, load at most $\xi$, and
its number of vertices is equal to $n$. Observe that $$n ~\ge~
N(\xi,h-1) ~\ge~ {h-1 \choose \xi} ~\ge~
\left(\frac{h-1}{\xi}\right)^{\xi},$$ implying that $$h ~\le~
n^{1/\xi} \cdot \xi + 1 ~=~ O(n^{1/\xi} \cdot \xi).$$
\QED
}

The next corollary employs Theorem \ref{htwolog} to deduce an analogous
result for general metric spaces.

\begin {corollary} \label{corhigh}
For  sufficiently large $n$ and $h$, $h \ge  2 \log n $, there
exists a binary spanning tree of $M$ that has depth at most $h$ and
lightness at most $2 \cdot \Psi$,
where $\Psi$ satisfies ($h = O(n^{1/\Psi} \cdot \Psi)$ and $\Psi =
O(\log n)$). Moreover, this binary tree can be constructed in time
$O(n^2)$.
\end {corollary}
{\bf Remark:} If $M$ is an Euclidean 2-dimensional metric space, the
running time can be further improved to $O(n \cdot \log n)$. This is
because the running time of this construction is dominated by the
running time of the subroutine for constructing MST, and an MST of
an Euclidean 2-dimensional metric space can be constructed in $O(n
\cdot \log n)$ time \cite{Epp96}. By the same considerations, for
Euclidean 3-dimensional spaces our algorithm can be implemented in a
randomized time of $O(n \cdot \log^{4/3}n)$, and more generally, for
dimension $d=3,4,\ldots$ it can be implemented in deterministic time
$O(n^{2-\frac{2}{\lceil d/2 \rceil +1} + \epsilon})$, for an
arbitrarily small $\epsilon > 0$ (cf.\ \cite{Epp96}, page 5). Even
better time bounds can be provided if one uses a
$(1+\epsilon)$-approximation MST instead of the exact MST for
Euclidean metric spaces (cf.\ \cite{Epp96}, page 6). \\
{
 \proof
 By Theorem \ref{htwolog} and Lemma \ref{vartogen},
for a sufficiently large $n$, and $h$, $h \ge  2 \log n $, there
exists a binary spanning tree $T$ of $M$ that has depth at most $h$
and load at most $\Psi$,
where $\Psi$ satisfies $h = O(n^{1/\Psi} \cdot \Psi)$.
By (\ref{xipsi}),
 the lightness $\Psi(T)$ is at most $2 \Psi$.
 Note that since the complete graph $G(M)$ induced by the metric
 $M$ contains at most $O(n^2)$ edges, its MST can be computed
 within $O(n^2)$ time (cf.\ \cite{CLRS90}, chapter 23). The in-order traversal, and the
 construction of the tree $T(\xi,h)$ can be performed in $O(n)$
 time in the straight-forward way. Hence the overall running time
 of the algorithm for computing an LLT with the specified
 properties is $O(n^2)$.
\QED
}

\begin{figure*}[htp]
\begin{center}
\begin{minipage}{\textwidth} 
\begin{center}
\setlength{\epsfxsize}{4.5in}
\epsfbox{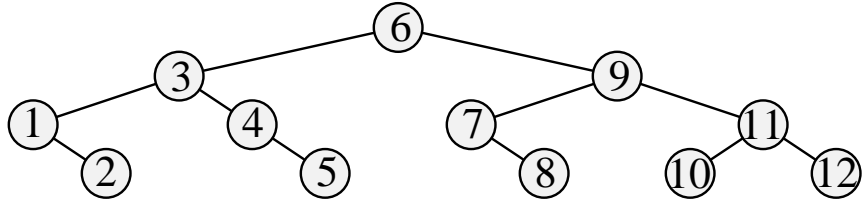}
\end{center}
\end{minipage}
\caption[]{
\label{fig:bal}
\sf
The full balanced binary tree for $n=12$ and $h=3$.
}
\end{center}
\end{figure*}

Finally, we present a simple construction for the range
$\log n \le h < 2 \log n$.
Consider a full\footnote{Strictly speaking, this tree is full when
$n = 2^{k}-1$, for integer $k \ge 1$. However, for simplicity of
presentation, we call it ``full" for other values of $n$ as well.}
balanced binary spanning tree $T_n$ of $\{1,2,\ldots,n\}$.
The root of $T_n$ is $\lceil n/2 \rceil$. Its left (respectively,
right) subtree is the full balanced binary tree constructed
recursively from the vertex set $\{1,\ldots,\lceil n/2 \rceil -1\}$
(resp., $\{\lceil n/2 \rceil +1,\ldots,n\}$). (See Figure
\ref{fig:bal} for an illustration.)

\begin {lemma} \label{tn}
Both the depth and the load of $T_n$ are no greater than $\log n$.
\end {lemma}
\proof
The proof is by induction on $n$. The base is trivial.
\\\emph{Induction Step:} We assume the correctness of the claim for all
smaller values of $n$, and prove it for $n$.
By the induction hypothesis, the depth of both the left and the
right subtrees of the root $\lceil n/2 \rceil$ is at most $\log
(n/2)$, implying that the depth $h(T_n)$ of $T_n$ is at most $\log
(n/2) + 1 = \log n$, as required.

To analyze the load $\xi(T_n)$ of the tree $T_n$, consider some edge
$e=(v_q,v_{q+1})$ of the path $L_{n}$. The two edges connecting $rt$
to its children contribute at most one unit to the load of $e$. In
addition, at most one among the subtrees of the root $\lceil n/2
\rceil$ covers the edge $e$. By induction, both of these subtrees
has load no greater than $\log (n/2)$. Consequently, the load of $e$
is at most $\log (n/2) + 1 = \log n$.
\QED

\begin {theorem} \label{corni}
For a sufficiently large $n$, and $h$, $ \log n \le h < 2 \log n$,
there exists a binary $\vartheta_n$-tree that has depth at most $h$
and load at most $\xi$,
where $\xi$ satisfies $h = O(n^{1/\xi} \cdot \xi)$. (In this case
$\xi= O(\log n)$.)
\end {theorem}
\proof
By Lemma \ref{tn}, for a sufficiently large $n$, and $h$, $\log n \le
h < 2 \log n$, the binary tree $T_n$ has depth at most $h$, and load
at most $\xi= \log n$, where $h = O(n^{1/\xi} \cdot \xi)$.
\QED

The next corollary employs Theorem \ref{corni} to deduce an analogous
result for general metrics.

\begin {corollary} \label{cormed}
For  sufficiently large $n$ and $h$, $ \log n \le h < 2 \log n$,
there exists a binary spanning tree of $M$ that has depth at most
$h$ and lightness at most $2 \cdot \Psi$,
where $\Psi$ satisfies $h = O(n^{1/\Psi} \cdot \Psi)$.
\end {corollary}
{
\proof
 By Theorem \ref{corni} and Lemma \ref{vartogen}, for a sufficiently
large $n$, and $h$, $ \log n \le h < 2 \log n$, there exists a binary
spanning tree of $M$ that has depth at most $h$ and load at most
$\Psi$,
where $\Psi$ satisfies $h = O(n^{1/\Psi} \cdot \Psi)$.
By
(\ref{xipsi}),
 the lightness is at most twice the load, and we are
done.
\QED } {\bf Remark:} By the same considerations as those of
Corollary \ref{corhigh}, this construction can be implemented within
time $O(n^2)$ in general metric spaces, and in time $O(n \cdot
\polylog(n))$ in Euclidean low-dimensional ones.

\subsection{Upper Bounds for Low Trees}
\label{sec:ublt} In this section we devise a construction of
$\vartheta_n$-trees with depth in the range of $h < \log n$.
Like the construction in Section
\ref{subhigh},
 this construction provides a tight \emph{up to constant factors}
 upper bound for the tradeoff between weight and depth.
 In addition, the maximum degree of the constructed trees is optimal
 as well. (It is $\lceil n^{1/h} \rceil$).

We devise a family of $d$-regular trees $\tilde T(d,h)$ with
${\displaystyle \tilde N(d,h) = \sum_{i=0}^h d^i}$ vertices, load
$\xi \le d \cdot h$, depth $h$, and $d \ge 2$. Since $d \ge 2$, it
holds that $\tilde N(d,h) = \frac{d^{h+1}-1}{d-1}$. The family
$\tilde T(d,h)$ is constructed recursively as follows. For $h=0$,
the tree $T_0 = \tilde T(d,0)$ is a singleton vertex, and so $\tilde
N(d,0) = 1$. For $h \ge 1$, the tree $\tilde T(d,h)$ is constructed
as follows. For each $i \in [d]$, let $T_i$ be a copy of $\tilde
T(d,h-1)$, and let $T_0 = \tilde T(d,0)$.
For every $i \in [\lceil d/2 \rceil]$, we add the tree $T_i$ as a
\emph{left} subtree of the root vertex $rt$ of $T_0$, and for each
$i \in [\lceil d/2 \rceil+1,d]$, we add the tree $T_i$ as a
\emph{right} subtree of $rt$.

\begin {lemma} \label{lotree}
The depth of the resulting tree  $\tilde T(d,h)$ with respect to the
vertex $rt$ is $h$, its load $\xi$ is at most $d \cdot h$, and its
size is equal to ${\displaystyle \sum_{i=0}^h d^i}$.
\end {lemma}
\proof
The proof is by induction on $h$. The base $h=0$ is trivial.
\\\emph{Induction Step:} We assume the correctness of the claim for all smaller values
of $h$, and prove it for $h$. By construction, the root $rt$ has $d$
children, each being the root of a tree $\tilde T(d,h-1)$. By the
induction hypothesis, $\tilde T(d,h-1)$ has depth $h$ and size
${\displaystyle \sum_{i=0}^{h-1} d^i}$. Obviously, the depth of
$\tilde T(d,h)$ is equal to $h$. To estimate the size $\tilde
N(d,h)$ of $\tilde T(d,h)$, note that $$\tilde N(d,h) = 1+ d \cdot
\tilde N (d,h-1) = 1+ d \cdot \left(\sum_{i=0}^{h-1} d^i\right) =
\sum_{i=0}^h d^i.$$

To analyze the load $\xi(\tilde T(d,h))$ of the tree $\tilde
T(d,h)$, consider some edge $e=(v_q,v_{q+1})$ of the path $L_{\tilde
n}$, where $\tilde n$ is the number of vertices in $\tilde T(d,h)$.
The $d$ edges connecting $rt$ to its children contribute altogether
at most $d$ units to the load of $e$. In addition, there exists at
most one index $i$, $i \in [h]$, such that the tree $T_i$ covers the
edge $e$. By the induction hypothesis, the load $\xi(T_i)$ of $T_i$ is no greater
than $d(h-1)$. Consequently, the load of $e$ is at most $d(h-1) + d
= d \cdot h$, and we are done.
\QED

\begin {theorem} \label{lot}
For any positive integers $n$ and $h$, $h < \log n$, there exists an
$\left \lceil n^{1/h}\right \rceil$-ary $\vartheta_n$-tree that has
depth at most $h$, and load at most $\left \lceil n^{1/h}\right
\rceil \cdot h$.
\end {theorem}
The case $h=0$ is trivial. \\For $h>0$, observe that $\tilde N(d,h)$
is monotone increasing with both $d$ and $h$. Given a pair of
positive integers $n$ and $h$, $1 \le h < \log n$, let $d$ be the
positive integer that satisfies $\tilde N(d-1,h) \le n < \tilde
N(d,h)$.

Consider the $d$-ary tree obtained from $\tilde T(d,h)$ by removing
$\tilde N(d,h) - n$ leaves from it, one after another. Clearly, the
depth, the load, and the maximum degree of the resulting tree are no
greater than those of the original tree $\tilde T(d,h)$. Moreover,
the size of the resulting tree is precisely $n$. Observe that
$$n ~\ge~ \tilde N(d-1,h)
~=~ \sum_{i=0}^h (d-1)^i ~>~ (d-1)^h,$$ implying that $d <
n^{1/h}+1$. It follows that $d \le \left \lceil n^{1/h} \right
\rceil$. Thus by Lemma \ref{lotree}, the resulting tree is a
spanning $\left \lceil n^{1/h} \right \rceil$-ary tree of size $n$
that has depth at most $h$, and load at most $\left \lceil n^{1/h}
\right \rceil \cdot h$, and we are done.

\QED

The next corollary is an extension of Theorem \ref{lot} to
general metric spaces.

\begin {corollary} \label{corlow}
For a sufficiently large $n$, and $h$, $h <  \log n $, there exists a
spanning $\left \lceil n^{1/h} \right \rceil$-ary tree of $M$ that
has depth at most $h$, and lightness at most $O(n^{1/h} \cdot h)$.
\end {corollary}
{
\proof
 By Theorem \ref{lot} and Lemma \ref{vartogen}, for a sufficiently
large $n$, and $h$, $h < \log n $, there exists a spanning $\left
\lceil n^{1/h} \right \rceil$-ary tree of $M$ that has depth at most
$h$ and load at most $\left \lceil n^{1/h} \right \rceil \cdot h$.
By (\ref{xipsi}),
 the lightness is at most twice the load, and we are done.
\QED } {\bf Remark:} By the same considerations as in Section
\ref{subhigh}, this construction can be implemented within time
$O(n^2)$ in general metric spaces, and in time $O(n \cdot
\polylog(n))$ in Euclidean low-dimensional ones. \\

Corollaries \ref{corhigh}, \ref{corlow} and \ref{cormed} imply the
following theorem.
\begin{theorem} \label{mainres}
For any sufficiently large integer $n$ and positive integer $h$,
and $n$-point metric space $M$, there exists a spanning tree of
$M$ of depth at most $h$ and lightness at most $O(\Psi)$, that
satisfies the following relationship. If $h \ge \log n$ then ($h
= O(\Psi \cdot n^{1/\Psi})$ and $\Psi = O(\log n)$). In the complementary range $h < \log n$,
$\Psi=O(h\cdot n^{1/h})$.
\\Moreover, this spanning tree is a binary one for $h \ge \log n$,
and it has the optimal maximum degree $\left \lceil n^{1/h} \right
\rceil$, for $h < \log n$.
\end {theorem}

\section {Shallow-Low-Light-Trees}
\label{sec:sllt}

In this section we extend our construction of low-light-trees and construct
shallow-low-light trees. Our argument in this section is closely related to
that of Awerbuch et al. \cite{ABP91}.

Consider a spanning tree $T = (V,E)$ of an $n$-point metric space $M$,
rooted at a root vertex $rt$, having depth $h(T)$ and weight
$\omega(T)$. We construct a spanning tree $S(T)$ of $M$ that has the
same depth and weight, up to constant factors, and that also
approximates the distances between the root vertex $rt$ and all
other vertices in $V$. (Thus, for a low-light tree $T$, $S(T)$ is a
shallow-low-light tree.)

 Let $L=(v_1,v_2,\ldots,v_n)$ be the sequence of vertices of $T$,
ordered according to an in-order traversal of $T$, starting at $rt$.
Fix a parameter $\theta$ to be a positive real number. The value of
$\theta$ determines the values of other parameters of the constructed tree.
\\We start with identifying a set of ``break-points" $\mathcal
B=\{B_1,B_2,\ldots,B_k\}$, $\mathcal B \subseteq V$. The
break-point $B_1$ is the vertex $v_1$. The break-point $B_{i}$ is
the first vertex in $L$ after $B_{i-1}$ such that
$$dist_T(B_{i-1},B_{i}) > \theta \cdot dist_M(rt,B_{i}).$$
Let $\mathcal S$ be the set of edges connecting the break-points with
$rt$ in $M$, namely, $\mathcal S = \{(rt,B_i) \vert B_i \in \mathcal
B\}$.
Let $\tilde G= (V,E \cup \mathcal S)$ be the graph obtained from $T$
by adding to it all edges in $\mathcal S$. \\Finally, we define
$S(T)$ to be the shortest-path-tree (henceforth, SPT) tree of
$\tilde G$ rooted at $rt$.

The following claim implies that the sum of distances in $T$, taken
over all pairs of consecutive break-points, is not too large.
\begin {claim} \label{diffb}
$\sum_{B_i \in \mathcal B} dist_T (B_{i-1},B_i) ~\le~ 2 \cdot
\omega(T).$
\end {claim}
\proof First, observe that any in-order traversal of $T$ visits each
edge exactly twice. Hence, since the vertices in $L$ are ordered
according to an in-order traversal of $T$, we have that
$\sum_{i=2}^n dist_T (v_{i-1},v_i) ~\le~ 2 \cdot \omega(T).$

Clearly, for any pair of vertices $v_i$ and $v_j$, $1 \le i \le j
\le n$, it holds that $dist_T(v_j,v_k) \le \sum_{i=j+1}^{k}
dist_T(v_{i-1},v_{i})$. It follows that $\sum_{i=2}^k dist_T
(B_{i-1},B_i) \le \sum_{i=2}^n dist_T (v_{i-1},v_i)$, and we are
done. \QED
%
The following three lemmas imply that the depth, the weight, and
the weighted diameter of the constructed tree $S(T)$ are not much
greater than those of the original tree $T$. In fact, if we set
$\theta$ to be a small constant, the three parameters do not
increase by more than a small constant factor. Moreover, not only
the weighted diameter does not grow too much, but, in fact, all
distances between the root vertex $rt$ and other vertices in the
graph are roughly the same in $T$ and in $S(T)$.
\begin {lemma} \label{firsty}
$h(S(T)) \le 1+2 \cdot (h(T)-1)$.
\end {lemma}
\proof A path from the root vertex $rt$ to a vertex $v$ in $S(T)$
may use an edge of $\mathcal S$ only as its first edge. If it does
so,
 its hop-length is at most $1+2(h(T)-1)$. Otherwise its hop-length is
 at most $h(T)$.
\QED

\begin {lemma} \label{secondy}
$\omega(S(T)) \le (1+2/\theta)\cdot \omega(T)$.
\end {lemma}
\proof Observe that \begin{equation} \label{wtbnd} \omega(S(T))
~\le~ \omega(\tilde G)
 ~=~
\omega(T) + \omega(\mathcal S) ~=~ \omega(T) + \sum_{B_i \in
\mathcal B} dist_M(rt,B_i).
\end {equation} By the choice of the break-points, for each $B_i \in
\mathcal B$,
$$dist_M(rt,B_i) ~<~ \frac{1}{\theta} \cdot dist_T (B_{i-1},B_i).$$
By Claim \ref{diffb},
$$\sum_{B_i \in \mathcal B} dist_T (B_{i-1},B_i) ~\le~ 2 \cdot \omega(T).$$
Therefore
$$\sum_{B_i \in \mathcal B} dist_M(rt,B_i) ~<~ \frac{1}{\theta} \cdot \sum_{B_i \in \mathcal B} dist_T (B_{i-1},B_i)
 ~\le~  \frac{2}{\theta} \cdot w(T),$$
 and thus (\ref{wtbnd}) implies that $\omega(S(T)) \le (1+2/\theta) \cdot \omega(T)$.
\QED

\begin {lemma} \label{thirdy}
For a vertex $v \in V$, it holds that
$$dist_{S(T)} (rt,v) \le (1+2\theta)\cdot dist_M(rt,v).$$
\end {lemma}
\proof Since $S(T)$ is an SPT with respect to $rt$ in $\tilde G$, it
suffices to bound $dist_{\tilde G} (rt,v)$. Let $i$ be the index
such that $v$ is located between $B_i$ and $B_{i+1}$ in $L$, $v \ne
B_{i+1}$. Since $B_i$ is a break-point, we have $dist_{\tilde G}
(rt,B_i) = dist_{M} (rt,B_i)$. Then
\begin{eqnarray*} dist_{\tilde
G}(rt,v) &\le& dist_{\tilde G} (rt,B_i) + dist_{\tilde G} (B_i,v) \\
&=& dist_M(rt,B_i) + dist_{\tilde G} (B_i,v) \\ &\le& dist_M(rt,B_i)
+ dist_T(B_i,v) \end{eqnarray*} Since $v$ was not selected as a
break-point, necessarily
\begin {eqnarray} \label{q1}
dist_T(B_i,v) \le \theta \cdot dist_M(rt,v).
\end {eqnarray}
Hence \begin {eqnarray} \label{q2} dist_{\tilde G} (rt,v) \le
dist_M(rt,B_i) + \theta \cdot dist_M(rt,v).
\end {eqnarray}
Observe that \begin {eqnarray*} dist_M(rt,B_i) &\le& dist_M(rt,v) +
dist_M(B_i,v) \\ &\le& dist_M(rt,v) + dist_T(B_i,v). \end
{eqnarray*} However, by (\ref{q1}), the right-hand side is at most
\begin {eqnarray*}
dist_M(rt,v) + \theta \cdot dist_M(rt,v) ~=~ (1+\theta) \cdot
dist_M(rt,v),
\end {eqnarray*}
which implies
\begin {eqnarray}  \label{q3}
dist_M(rt,B_i) &\le& (1+\theta) \cdot dist_M(rt,v).
\end {eqnarray}
Plugging (\ref{q3}) in (\ref{q2}), we obtain $dist_{\tilde G} (rt,v)
\le (1+2 \theta) \cdot dist_M(rt,v)$.
\QED

Set $\epsilon = \frac{\theta}{2}$. Lemmas \ref{firsty},
\ref{secondy} and \ref{thirdy} imply the following theorem.
\begin {theorem} \label{slltt}
Given a rooted spanning LLT $(T,rt)$ of a metric space $M$ that has
depth at most $h$ and weight at most $W$, the tree $S(T)$ is a
spanning SLLT of $M$ that has depth at most $2h-1$ and weight $O(W
\cdot (\epsilon^{-1}))$. In addition, for every point $v$,
$dist_{S(T)}(rt,v) \le (1+ \epsilon) \cdot dist_M (rt,v)$.
\end {theorem}

Theorems \ref{mainres} and \ref{slltt} imply the following corollary, which is the
main result of this section.
\begin {corollary}
For a sufficiently large integer $n$, a positive integer $h$, a
positive real $\epsilon > 0$,  an $n$-point metric space $M$, and a
designated root point $rt$, there exists a spanning tree $T$ of
$M$ rooted at $rt$ with hop-radius at most $O(h)$ and lightness
at most $O(\Psi \cdot (\epsilon^{-1}))$, such that ($h = O(\Psi
\cdot n^{1/\Psi})$ and $\Psi = O(\log n)$) whenever $h \ge \log n$, and
$\Psi = O(h \cdot n^{1/h})$ whenever $h < \log n$. Moreover, for every
point $v \in M$, the weighted distance between the root $rt$ and
$v$ in $T$ is greater by at most a factor of $(1+\epsilon)$ than
the weighted distance between them in $M$.
\end {corollary}
This construction can also be implemented within $O(n^2)$ time in
general metric spaces, and within $O(n \cdot \polylog(n))$ time in
Euclidean low-dimensional ones. The only ingredient of this
construction that is not present in the constructions of Sections
\ref{sec:ub} and \ref{sec:ublt} is the construction of the
shortest-path-tree for $\tilde G$. We observe, however, that $\tilde
G$ has only a linear number of edges, and thus, this step requires
only $O(n \cdot \log n)$ additional time.

\begin{figure*}[htp]
\begin{center}
\begin{minipage}{\textwidth} 
\begin{center}
\setlength{\epsfxsize}{3.5in}
\epsfbox{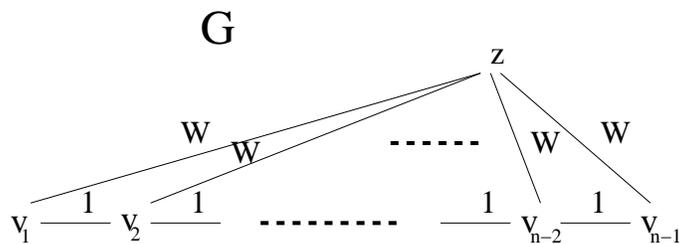}
\end{center}
\end{minipage}
\caption[]{
\label{huge}
\sf
The graph $G=(V,E,\omega)$.
}
\end{center}
\end{figure*}

Finally, we show that there are graphs for which any spanning tree
has either huge hop-diameter or huge weight. Specifically, consider
the graph $G=(V,E,\omega)$ formed as  union of the $(n-1)$-vertex
path $P_{n-1} = (v_1,v_2,\ldots,v_{n-1})$ with the star $S =
\{(z,v_i) \vert i \in [n-1]\}$. All edges of the path have unit
weight, and all edges of $S$ have weight $W$, for some large integer
$W$, $n \ll W$. (See Figure \ref{huge} for an illustration.) Note
that $G$ is a \emph{metric graph}, that is, for every edge $(u,w)
\in E$, $\omega(u,w) = dist_G(u,w)$.  It is easy to verify that any
spanning tree $T$ of $G$ that has weight at most $q \cdot W$, for
some integer parameter $q$, $q \ge 1$, has hop-diameter
$\Omega(n/q)$. Symmetrically, for an integer parameter $D$, $D \ge
1$, any spanning tree $T$ of $G$ that has hop-diameter at most $D$
contains at least $\Omega(n/D)$ edges of weight $W$. Since the
weight of the MST of $G$ is only slightly greater than $W$, it
follows that the lightness of $T$ is at least $\Omega(n/D)$.

Consequently, for any general construction of LLTs for
\emph{graphs}, $\Psi \cdot \Lambda = \Omega(n)$. As we have shown,
for \emph{metric spaces} the situation is drastically better,
and, in particular, one can have both $\Psi$ and $\Lambda$ no
greater than $O(\log n)$.

\section*{Acknowledgements}

  The second-named author thanks Michael Segal and Hanan Shpungin for
  approaching him with a problem in the area of wireless networks that
  is related to the problem of constructing LLTs.
Correspondence with them triggered this research.


\newpage

\def\thepage{}

\clearpage
\pagenumbering{roman}
\appendix
\centerline{\LARGE\bf Appendix}
\section {Properties of the Binomial Coefficients} \label{app}
In this section we
present a number of useful properties of the binomial coefficients.

The following two statements are well-known \cite{GKO94}.
\begin {fact} [Pascal's 2nd identity] \label{fact23}
For any non-negative integers $h$ and $i$, such that $i \le h$,
$$\sum_{k=i}^h {k \choose i} ~=~ {h+1 \choose i+1}.$$
\end {fact}

\begin {fact} [Pascal's 7th identity] \label{fact2}
For any non-negative integers $n$ and $k$, such that $k+1 \le n$,
$${n \choose k+1} = \frac{n-k}{k+1}\cdot {n \choose k}.$$
\end {fact}

The next lemma shows that the sequence of binomial coefficients
$\mathcal B = \{{n \choose i} ~\vert~ i=1,2,\ldots,n\}$ grows
exponentially with $i$ as long as $i \le \lfloor \frac{n}{4}
\rfloor$.
\begin {lemma} \label{lemm1}
For any non-negative integers $n$ and $k$, such that $k \le
\lfloor \frac{n}{4} \rfloor$,
$$\sum_{i=0}^{k} {n \choose i} < \frac{3}{2} \cdot {n \choose k}.$$
\end {lemma}
\proof
By Fact \ref{fact2}, for any $0 \le i \le k-1$, it holds that
$${n \choose i+1} ~=~ \frac{n-i}{i+1}\cdot {n \choose i} ~\ge~
 \frac{n-\lfloor \frac{n}{4} \rfloor +1}{\lfloor \frac{n}{4} \rfloor}\cdot {n \choose i} ~>~ 3\cdot {n \choose i}.$$
 Hence, $$\sum_{i=0}^{k} {n \choose i} ~\le~ {n \choose k} \cdot
 \sum_{i=0}^k \left(\frac{1}{3}\right)^i ~<~  \frac{3}{2} \cdot {n \choose k}.$$
\QED

\end{document}